%% file: ms.tex
\documentclass[journal]{vgtc}                
\ifpdf
  \pdfoutput=1\relax                   
  \usepackage{graphicx}                
  \DeclareGraphicsExtensions{.pdf,.png,.jpg,.jpeg} 
\else
  \usepackage{graphicx}                
  \DeclareGraphicsExtensions{.eps}     
\fi%

\graphicspath{
{4_TestFunctions/4_1_Local_Attributes/pictures/}
{4_TestFunctions/4_2_Global_Attributes/pictures/}
{4_TestFunctions/4_3_RealWorld/pictures/}
{5_TestEvaluation/pictures/}
{6_ApplicationCase/pictures/}
} 

\usepackage{microtype}                 
\PassOptionsToPackage{warn}{textcomp}  
\usepackage{textcomp}                  
\usepackage{mathptmx}                  
\usepackage{times}                     
\usepackage{cite}                      
\usepackage{tabu}                      
\usepackage{booktabs}                  

\onlineid{0}

\vgtccategory{Research}
\vgtcpapertype{system}

\title{A Testing Environment for Continuous Colormaps}

\author{P. Nardini,
        M. Chen, \textit{Member, IEEE},
        R. Bujack,
        M. B\"ottinger,
        and G. Scheuermann, \textit{Member, IEEE}
        }
\authorfooter{
\item
  Pascal Nardini is with Institute of Computer Science, University of Leipzig, Germany. E-mail: nardini@informatik.uni-leipzig.de.
\item
  Min Chen is with Department of Engineering Science, University of Oxford, UK. E-mail: min.chen@oerc.ox.ac.uk. 
\item
  Roxana Bujack is with Data Science at Scale Team, Los Alamos National Laboratory, Los Alamos, NM, USA. E-mail:bujack@lanl.gov.
\item 
  Michael B\"ottinger is with German Climate Computing Center (DKRZ), Hamburg, Germany. E-mail:boettinger@dkrz.de. 
\item 
  Gerik Scheuermann is with Institute of Computer Science, University of Leipzig, Germany. E-mail:scheuermann@informatik.uni-leipzig.de. 
}

\shortauthortitle{Biv \MakeLowercase{\textit{et al.}}: A Testing Environment for Continuous Colormaps}

\abstract{
Many computer science disciplines (e.g., combinatorial optimization, natural language processing, and information retrieval) use standard or established test suites for evaluating algorithms.
In visualization, similar approaches have been adopted in some areas (e.g., volume visualization), while user testimonies and empirical studies have been the dominant means of evaluation in most other areas, such as designing colormaps.
In this paper, we propose to establish a test suite for evaluating the design of colormaps.
With such a suite, the users can observe the effects when different continuous colormaps are applied to planar scalar fields that may exhibit various characteristic features, such as jumps, local extrema, ridge or valley lines, different distributions of scalar values, different gradients, different signal frequencies, different levels of noise, and so on.
The suite also includes an expansible collection of real-world data sets including the most popular data for colormap testing in the visualization literature.
The test suite has been integrated into a web-based application for creating continuous colormaps (https://ccctool.com/), facilitating close inter-operation between design and evaluation processes.
This new facility complements traditional evaluation methods such as user testimonies and empirical studies.
} 


\CCScatlist{ 
 \CCScat{K.6.1}{Management of Computing and Information Systems}%
{Project and People Management}{Life Cycle};
 \CCScat{K.7.m}{The Computing Profession}{Miscellaneous}{Ethics}
}

\teaser{
  \centering
  \includegraphics[width=0.99\linewidth]{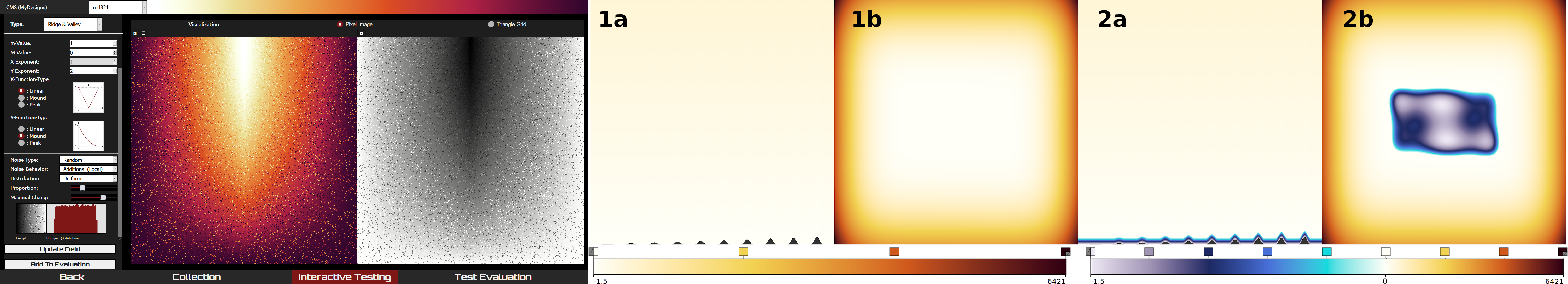}
  \caption{
    \textbf{Left}: Screenshot of our new interactive test suite implemented in the CCC-Tool.
    \textbf{Right}: Two visualizations show at \textbf{b} the valley shaped Six-Hump Camel Function \cite{optimization1999handbook} for the area $[-5,5]\times[-5,5]$ and at \textbf{a} the $Little Bit$ test function (\autoref{subsubsection:littleBitVariations}).
    Special features of the Six-Hump Camel function are irregularly shaped troughs in the center of the area that are more than two orders smaller than the data range difference between center and boundary.
    The uniformly spaced colormap \textbf{1} cannot reveal the troughs, while the last nonlinear colormap \textbf{2} clearly shows the topological structure of the irregularities in the center of the domain.
  }
  \label{fig:camelExample}
}

\vgtcinsertpkg

\usepackage{amssymb}
\usepackage{amsmath}

\AtBeginDocument{%
  \let\mathbb\relax
  \DeclareMathAlphabet\PazoBB{U}{fplmbb}{m}{n}
  \newcommand{\mathbb}{\PazoBB}
}

\begin{document}


\firstsection{Introduction\label{section:introduction}}

\maketitle

\input{1_0_Introduction.tex}
\input{2_0_RelatedWorks.tex}

\input{3_0_Motivation_Design.tex}
\input{4_0_Testing_Functions.tex}

\input{5_0_TestEvaluation.tex}
\input{6_0_applicationCase.tex}

\input{7_0_Conclustion.tex}


\bibliographystyle{abbrv-doi}

\bibliography{references,references_tools,references_optimization,references_math,references_analysis_and_evaluation}

\end{document}

%% file: 1_0_Introduction.tex
%


In many fields in computer science, algorithms are commonly evaluated using open libraries of predefined tests.
A typical example is the field of combinatorial optimization with ACM GECCO~\cite{GECCO:2018} as its leading conference.
New algorithms are often tested against standard problems (e.g., the traveling salesman problem or the quadratic assignment problem), using established libraries (e.g., SPlib~\cite{Reinelt:1991} or QAPlib~\cite{Burkard:1997}).
These libraries contain test data sets with increasing difficulty, each typically focusing on different (known) challenges in combinatorial optimization.
Similarly, the natural language processing (NLP) community frequently uses test data sets in the Semantic Evaluation suite \cite{SemEval:2020:web}.
In the past, algorithmic developments in NLP benefited substantially from the TREC collection of test data \cite{TREC:2020:web}.

Algorithmic development has been a core component of scientific visualization.
There were earlier attempts to create an open collection of test data, such as in the area of volume visualization, and no many collections are still remaining (e.g., Klacansky:2020:web).
Whilst there are data sets from the SciVis contests, the evaluation of an algorithm or the comparison of different algorithms is usually conducted with proprietary data sets chosen by the evaluator (who is often also the algorithm developer).
In some cases, the evaluation or comparison is accompanied by a usability study involving domain experts.
A colormap implicitly encodes an algorithmic transformation from a data set to its visual representation.
Hence the evaluation of a transformation is primarily about the colormap concerned.
In this paper, we report a test suite for evaluating continuous colormaps, in the spirit of the open approach seen in other fields as well as in the earlier decades of scientific visualization.
This does not mean to abandon usability studies, but the test suite can efficiently answer many questions and help improve colormap designs before or after a usability study.
The test suite will also allow the VIS community to collect test data representing different challenges to colormap designs or different applications as well as helping identifying guidelines to use or pitfalls to avoid.

Color mapping is probably the mostcommonly used method for transforming data variables into visual variables.
Color heatmaps are ubiquitous in all scientific applications featuring captured or simulated data with a spatial context (e.g., geographical data, imagery data, simulation results underpinned by some geometry, and so on).
Every domain expert dealing with 2D scalar fields knows how to observe color heatmap.
Color mapping is also an integral component in almost all complex rendering techniques for visualizing height-, scalar-, vector-, and tensor-fields.
One would expect to find some color mapping methods in any free and commercial visualization software system~\cite{Stalling::2005::Amira, Ahrens:2005:ParaView, Childs:2012:VisIt}.
 
Selecting a good (continuous) colormap has been an enduring topic in visualization for many years (e.g., \cite{rogowitz1998data,luo2001development}).
There are several metrics for measuring the quality of colormaps (e.g., perceived color differences~\cite{,Bujack:2018:TVCG}) and tools for creating colormaps (e.g., \cite{nardini2019making}).
However, there is yet a test suite for colormap designers to try out some design options in order to investigate how they may reveal patterns in different test data sets.
Although one may test a colormap using an existing application data set, this is usually not adequate enough because given a context, a color map should not only reveal important patterns in known data sets, but also do the same for new data sets that may arrive in the future.
A test suite can enable designers to speculate on different ``what-if'' scenarios, and examine how a colormap may react.
%
This paper presents a test suite designed to serve the above purpose.
The test suite consists of three parts.
The first part contains a set of test functions that provide a comprehensive set of potential \emph{local properties} of 2D scalar functions.
For example, some functions may feature ``jumps'' at different value levels and with different increments or decrements, gradients of different scales starting from different levels, critical points, ridges, and different frequencies.
In \autoref{section:testingFunctions}, we give a mathematical reasoning behind the selection of these functions.

The second part of the test suite contains functions to model a set of \emph{global properties}.
The designer of a colormap often encounters challenging data sets with complex variations of data ranges in different regions.
For example, a multi-band colormap may suit a region with large but smooth jumps, while a slow-paced sequential colormap may suit a region with many small but volatile jumps.
Hence test functions that simulate such complex variations challenge colormap designers as well as colormap optimization algorithms.
We use some well-known sample functions from the (continuous) optimization literature for this purpose.
Complex variations can also be caused by some local properties alone, such as different signal-to-noise levels.

In addition to the two categories of test functions, the third part provides \emph{real-world data sets} collected from different applications.
We expect that this part of the test suite will continue to expend.
 
%
By integrating the test suite in our open-access CCC-Tool, which enables users to create, edit and analyze colormaps~\cite{nardini2019making}, we now provide a complete design and test environment for developing and testing of application-specific colormaps and for supporting  all types of visualization applications and domain-specific data sets. 

%% file: 2_0_RelatedWorks.tex
%

\section{Related Work} \label{section:relatedwork}

\subsection{Color Mapping}
Continuous color mapping (also heat mapping) refers to the association of a color to a scalar value over a domain and can be considered the most popular visualization technique for two-dimensional scalar data. 

There are many heuristic rules for designing a good colormap, which have been applied mostly intuitively for hundreds of years~\cite{Silva::2017::UseColorInVis, silva2011using, zhou2016survey}.
The most important ones are order, high discriminative power, uniformity, and smoothness~\cite{Bujack:2018:TVCG}. 

While some colormaps have been designed to sever as default colormaps for many data sets and can perform reasonably well in terms of rule compliance~\cite{moreland2009diverging}, many colormaps are purposely-designed according to application-specific requirements such as the shape of the data, the audience, the display, or the visualization goal~\cite{sloan1979color, bergman1995rule, rheingans2000task, borland2011collaboration}. 
The number of possible colormap configurations and the body of related work on this topic are huge~\cite{ware1988color, bergman1995rule, rogowitz1996not, rheingans2000task, tominski2008task}. 

An effort has been made to measure the quality of colormaps with respect to these rules quantitatively~\cite{tajima1983uniform, robertson1986generation, levkowitz1992design, moreland2009diverging, Bujack:2018:TVCG} or experimentally~\cite{ware1988color, rogowitz1999trajectories, kalvin2000building, ware2017uniformity}, in order to develop theories and algorithms that can help automate the generation, evaluation, and improvement of colormaps.
Although such theories and algorithms are usually general enough to be application-independent, the design of colormaps in many practical applications can only be effective if one includes some application-specific semantics in the design, such as key values, critical ranges, non-linearity, probability distribution of certain values or certain spatial relationships among values, and so on.
Supporting such application-specific design effort is the goal of this work.

\subsection{Colormap Test Data}
So far, there is no test suite for colormaps.
However, the literature has provided various examples where some data sets were used for comparing color maps and demonstrating color mapping algorithms.

Sloan and Brown \cite{sloan1979color} suggest treating a colormap as a path through a color space and stress that the optimal choice of a colormap depends on the task, human perception, and display. They showcase their findings with $x$-ray and satellite images.
Wainer and Francolini \cite{wainer1980empirical} point out the importance of order in a visual variable using statistical information on maps.
Pizer \cite{pizer1981intensity, pizer1982concepts} stresses the importance of uniformity in a colormap, of which the curve of just noticeable differences (JNDs) is constant and in a natural order.
The uniformity can be achieved by increasing monotonically in brightness or each of the RGB components, such that the order of their intensities does not change throughout the colormap.
Tajima \cite{tajima1983uniform} uses colormaps with regular color differences in a perceptually uniform colorspace to achieve perceptually uniform color mapping of satellite images.
Levkowitz and Herman \cite{levkowitz1992design} suggest an algorithm for creating colormaps that produces maximal color differences while satisfying monotonicity in RGB, hue, saturation, and brightness. They test them with medical data from CT scans.
Bernard et al. \cite{bernard2015survey} suggest definitions of colormap properties and build relations to mathematical criteria for their assessment and map them to different tasks independent from data properties in the context of bivariate colormaps. 
They test the criteria on analytical data that has different shapes (e.g., different gradients and spherical surfaces).

Pizer states that the qualitative task is more important in color mapping applications, because quantitative tasks can be better performed using contours or by explicitly displaying the value when the user hovers over a data point with the mouse.
He uses medical images, including CT scans and digital subtraction fluorography, as example data sets.
Ware \cite{ware1988color} also distinguishes qualitative and quantitative tasks.
He agrees that the qualitative characteristics are more important and explicitly mentions tables as a suitable means for the visualization of quantitative values.
On the one hand, he finds that monotonic change in luminance is important to see the overall form (qualitative) of his analytic test data consisting of linear gradients, ridges, convexity, concavity, saddles, discontinuities, and cusps.
On the other hand, his experiments show that when a colormap consists of only one completely monotonic path in a single perceptual channel, the quantitative task is error-prone if one tries to read the exact data values based on the visualization.

Rogowitz, et al.~\cite{rogowitz1992task, rogowitz1994using, bergman1995rule, rogowitz1996not, rogowitz1998data, kalvin2000building, rogowitz2001blair} distinguish different tasks (isomorphic, segmentation, and highlighting), data types, (nominal, ordinal, interval, and ratio), and spatial frequency (low, high), recommending colormap properties for each combination.
They perform experiments on the visual perception of contrast in colormaps using Gaussian or Gabor targets of varying strength superimposed on linear gradients of common colormaps~\cite{rogowitz1999trajectories, kalvin2000building}.
Rogowitz et al. use a huge variety of data through their extensive experiments, for example, text\cite{rogowitz1992task}, MRI scans of the human head \cite{rogowitz1998data, rogowitz1998data}, weather data showing clouds or ozone distribution \cite{rogowitz1998data, rogowitz1998data}, vector field data from a simulation of the earth's magnetic field or jet flows \cite{rogowitz1998data, bergman1995rule}, measurements from remote sensing \cite{rogowitz1996not}, cartographic height data \cite{rogowitz1998data}, analytic data covering a broad spectrum of frequencies such as planar wave patterns \cite{rogowitz1996not}, linear gradients distorted by a Gaussian or Gabor target of increasing magnitude \cite{rogowitz1999trajectories, kalvin2000building}, the luminance of a human face photograph \cite{rogowitz2001blair}, and so on.
Their work demonstrates the diversity in the application field of color mapping and how important it is for a colormap to encode application-specific semantics.
Zhang and Montag \cite{zhang2006perceptual} evaluate the quality of colormaps designed in a uniform color space with a user study using a CAT scan and scientific measurements such as remote sensing and topographic height data.
Gresh \cite{gresh2010self} measures the JND between colors in a colormap, using cartographic height data.
Ware et al.~\cite{ware2017uniformity} generate stimuli for experiments on colormap uniformity by superimposing vertical strips of Gabor filters of different spatial extent over popular colormaps with magnitudes ranging from nonexistence on the top to very strong contrast on the bottom.
The users' task is to pick the location where they could first perceive the distortion.

Light and Bartlein \cite{light2004end} warn of using the rainbow colormap, showing that it is highly confusing for color vision impaired users at the example of temperature data covering North and South America.
Borland \cite{borland2007rainbow} also criticizes the rainbow colormap for its lack of order.
He compares different colormaps based on analytic test data that features a spectrum of changing frequencies, different surface shapes, and gradients.
Kindlmann et al. \cite{kindlmann2002face} suggest a method to evaluate users' perception of luminance using a photograph of a human face.
Schulze-Wollgast et al. \cite{schulze2005enhancing} focus on the task of comparing data using statistical information on maps.
Tominski et al. \cite{tominski2008task} also stress that the characteristics of the data, tasks, goals, user, and output device need to be taken into account.
They introduce their task-color-cube, which gives recommendations for the different cases.
They use cartographic data to demonstrate their findings.
Wang \cite{Wang:2008} chooses color for illustrative visualizations using medical data and measurements of transmission electron microscopy (TEM), analytic jumps, and mixing of rectangles.
Zeileis et al. \cite{zeileis2009escaping} provide code to generate color palettes in the cylindrical coordinates of CIEUV and showcase results using geyser eruption data of Old Faithful and cartographic data. 
Moreland \cite{moreland2009diverging} presents an algorithm that generates diverging colormaps that have a long path through CIELAB without sudden non-smooth bends.
His red-blue diverging colormap is the current default in ParaView \cite{Ahrens:2005:ParaView}.
He tests different colormaps with data representing a spectrum of frequencies and gradients partly distorted by noise.
He also stresses the importance of testing on 3D surfaces where shading and color mapping compete, e.g., the density on the surface of objects in flow simulation data or on 3D renderings of cartographic height data. 
Borland \cite{borland2011collaboration} collaborates with an application scientist working on urban airflow.
They suggest combining existing colormaps to design domain-specific ones, and in case of doubt stick with the black-body radiation map. %
They sacrifice traditional rules (e.g., order) to satisfy the needs (huge discriminative power) of the application.
Eisemann et al. \cite{eisemann2011data} separate the adaption of the histogram of the data from the color mapping task, introducing an interactive pre-colormapping transformation for statistical information on maps.
Thompson et al. \cite{thompson2013provably} suggest applying special colors outside the usual gradient of the colormap to dominantly-occurring values, which are ``prominent'' values occurring with high frequency.
Their test data includes the analytic Mandelbrot fractal and flow simulation results, which are partly provided as examples in ParaView. 
Brewer \cite{brewer1994color,Brewer:2004:designing} provides an online tool to choose carefully designed discrete colormaps.
This is perhaps the most widely used tool for discrete colormaps.
Mittelst\"adt et al. \cite{mittelstaedt2014methods,mittelstaedt2015colorcat} present a tool that helps to find a suitable colormap for different task combinations.
They showcase their findings with analytical data, like gradients and jumps, and real-world maps.
Samsel et al. \cite{Samsel:2015:CHI, samsel2017envir} provide intuitive colormaps designed by an artist to visualize ocean simulations and scientific measurements in the environmental sciences.
Fang et al.~\cite{Fang:2017:TVCG} present an optimization tool for categorical colormaps, and use the tool to improve the colormap of the London underground map and that for seismological data visualization.
Nardini et al.~\cite{nardini2019making} provides an online tool, the \texttt{CCC-Tool}, for creating, editing, and analyzing continuous colormaps, demonstrating its uses with captured hurricane data, simulated ocean temperature data, and results of simulating ancient water formation.

\begin{table}[t]
\vspace{0mm}
\caption{The most popular test data for colormap testing in the visualization literature.\label{t:related}}
\centering
\begin{tabular}{@{\hspace{4mm}}r@{\hspace{4mm}}l@{\hspace{4mm}}}
\hline
analytic data & 
\cite{ware1988color, rogowitz1996not, rogowitz1999trajectories, kalvin2000building, borland2007rainbow, Wang:2008, moreland2009diverging, thompson2013provably, mittelstaedt2014methods}\\
& \cite{mittelstaedt2015colorcat, bernard2015survey, ware2017uniformity}\\
statistics and maps& \cite{sloan1979color, brewer1994color, Brewer:2004:designing, schulze2005enhancing, tominski2008task, zeileis2009escaping, eisemann2011data, mittelstaedt2014methods, Fang:2017:TVCG} \\
medical imaging &  \cite{sloan1979color, pizer1981intensity, pizer1982concepts, levkowitz1992design, rogowitz1998data, zhang2006perceptual, Wang:2008}  \\
scientific measurements & \cite{rogowitz1996not, rogowitz1998data, light2004end, zhang2006perceptual, moreland2009diverging, gresh2010self, zeileis2009escaping, samsel2017envir, Fang:2017:TVCG, nardini2019making} \\
scientific simulations & \cite{bergman1995rule, rogowitz1998data, moreland2009diverging, thompson2013provably, borland2011collaboration, Samsel:2015:CHI, samsel2017envir, nardini2019making}  \\
photographs & \cite{sloan1979color, tajima1983uniform,rogowitz2001blair, kindlmann2002face}   \\
\hline
\end{tabular}
\vspace{-4mm}
\end{table}

All in all, we found that the most popular way of evaluating the quality of colormaps in the literature is the use of specifically designed analytic data like gradients, ridges, different surface shapes, fractals, jumps, or different frequencies, because these synthetic data sets help to identify specific properties of the colormaps.
The second most common use is cartographic maps, which reflects the historical use of color mapping.
Furthermore, it is also common to use data in typical applications of scientific visualization as test data, e.g., fluid simulations (wind, ocean, turbulence), scientific measurements (weather, clouds, chemical concentration, temperature, elevation data), and medical imaging (x-ray, CT scan, digital subtraction fluorography, transmission electron microscopy).
A summary can be found in \autoref{t:related}.
We have carefully designed our colormap test suite according to these findings, not only providing an extensive selection of expressive analytic data, but also containing real-world data from different scientific applications.

%% file: 3_0_Motivation_Design.tex
%

\section{Motivation} \label{section:motivationAndDesign}
In several fields of computer science, the use of established test suites for evaluating techniques is standard or commonplace.
The motivation for this paper is to introduce such a test suite to scientific visualization. 
So far, user testimonies and empirical studies have been the dominant means of evaluation in the literature.
With this work, we would like to initiate the development of an open resource that the community can use to conduct extensive and rigorous tests on various colormap designs.
We also anticipate that the community will contribute new tests to this resource continuously, including but not limited to tests for colormaps used in vector- or tensor-field visualization.
Such a test suite can also provide user-centered evaluation methods with stimuli and case studies, while stimulating new hypotheses to be investigated using perceptual and cognitive experiments. 

%
The development of testing functions in this paper deals with the common features that pose challenges in scalar analysis, such as jumps, local extrema, ridge or valley lines, different distributions of scalar values, different gradients, different signal frequencies, different levels of noise, and so on.
This scope should be extended in future work progressively with more and more complex or specialized cases.

The main design goal of our test suite is to provide a set of intuitive functions, each of which deals with one particular challenge at a time.
They should be easy to interpret and to customize by experts as well as non-expert users.
This aspired simplicity in design can be exploited in future work to facilitate automatic production of test reports or automatic optimization of colormaps with respect to a selection of tests.

At present, this initial development should provides a set of test functions simulating a variety of planar scalar fields with different characteristic features.
It should enable the users to observe the effects when different continuous colormaps are applied to scalar fields that have the characteristic features similar to those featured in an application.
In many situations, the users may anticipate certain features in data sets that are yet to arrive, and would like to ensure that the color mapping can reveal such features effectively when the data arrives.
Finding and defining a suitable testing function is usually easier than manually creating a data set.
Especially, unlike a synthetic data set, a test function is normally resolution-independent and is accompanied by some parameters for customization.

In addition, the test suite should provide users with data sets that come from real-world applications, possibly with some modification wherever appropriate. Such an application-focused collection can be compiled from the most popular data for colormap testing in the visualization literature.
Since both the collection of test functions and that of real-world data sets are extensible, the field of visualization may soon see a powerful test suite for evaluating colormap design.
This is desirably in line with other computer science disciplines.

%% file: 4_0_Testing_Functions.tex
%

\section{Test Suite} \label{section:testingFunctions}
The first design goal of our test functions is to allow intuitive interpretation of colormap properties by users.
This requires each test function to have an easily-understandable behavior, and to have a clear mathematical description that can be reproduced consistently across different implementation.
The second design goal is to build the collection of test functions on the existing analytic examples in the literature surveyed in \autoref{section:relatedwork} to ensure that the existing experiments can be repeated and compared with new experiments.
The third design goal is to help users to find the test suite and to conduct tests easily.
Hence we integrate the test suite with the \texttt{CCC-Tool}, allowing users to conduct tests immediately after creating or editing a colormap specification.

As mentioned in \autoref{section:introduction}, our test suite has three parts: local tests, global tests, and a set of application data sets.
The local tests are mostly based on analytic examples in the literature and are defined with considerations from calculus to cover most local properties of scalar functions.
The global tests feature analytic properties of scalar functions that are not local, such as signal to noise ratio, global topological properties, different levels of variation, etc.
Finally, the application-specific data sets reflect the well-documented fact that colormaps should also be evaluated using real-world data sets.

The mathematical notions in this section use a few common parameters. The user-defined parameters, $r, R$, set the test function range, with $r, R \in \mathbb{R} \land r \neq R$. $R$ and $r$ determine the minimum $m$ and maximum $M$ of the test function with $m < M \in \mathbb{R}$. With $b \in \mathbb{N}$ the user can select an exponent that describes the polynomial order.
For functions with enumerated cases, the user can select a specific option $T$.
 
  \begin{figure}[t] 
  	\centering
  	\includegraphics[width=0.99\linewidth]{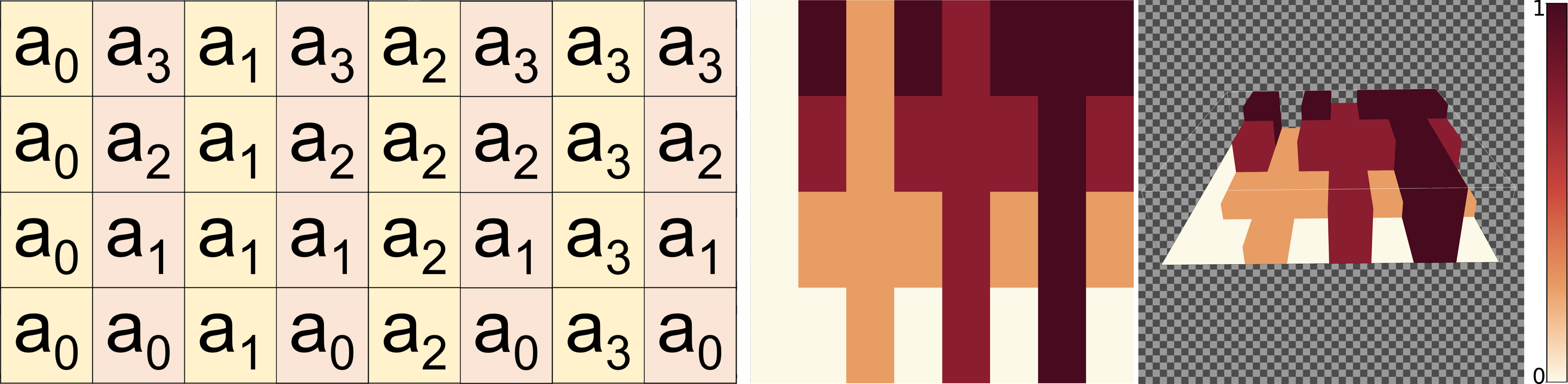}
  	\caption{ \label{fig:stepExample}
  	  \textbf{Left:} The table shows the structure of the neighborhood with four elements ($A=\{a_0,a_1,a_2,a_3\}$). The odd indexed columns (yellow) always include the same value, increasing from the first to the last column. The even indexed columns (orange) contain the whole set of test values in increasing order. 
  	  \textbf{Right:} Neighborhood variation test with $A=\{0.0,0.25,0.75,1.0\}$ for the colormap displayed below including a three-dimensional version encoding the values through height.
  	  }
  \end{figure}
  
  \begin{figure}[t] 
  	\centering
  	\includegraphics[width=0.8\linewidth]{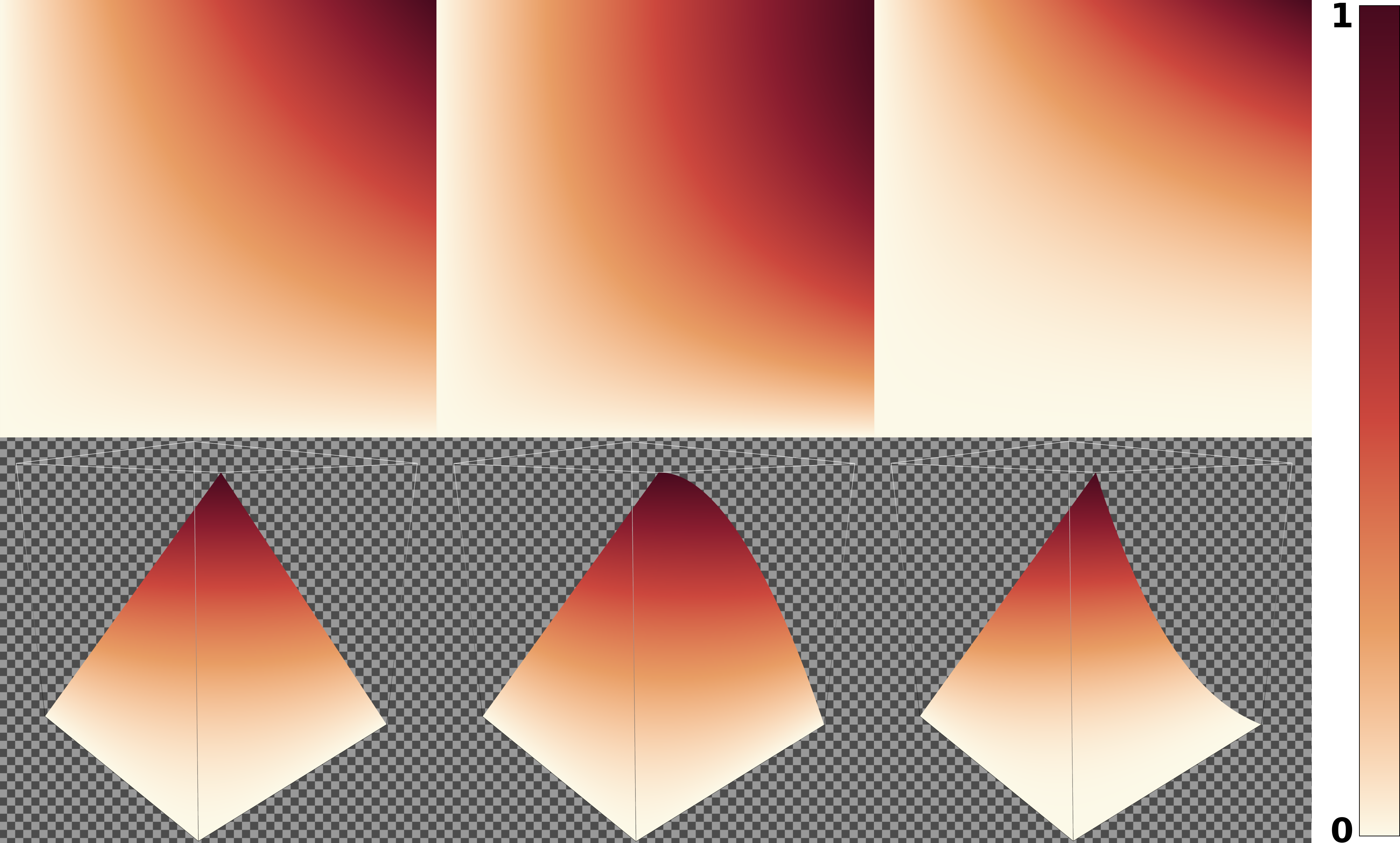}
  	\caption{ \label{fig:gradientExample} Three gradient tests, with $r=0$, $R=1.0$.
      	\textbf{Top Row}: Color mapping visualizations of the \texttt{Gradient Variation} function for the types $linear$, $convex$, and $concave$ with $T_x=T_y$ and $b=1$ for the first type and $b=2$ for the other types.
      	\textbf{Bottom Row}: 3D height-map visualizations of the three gradient tests.
  	}
  \end{figure}
  
  \begin{figure}[t] 
  	\centering
  	\includegraphics[width=0.8\linewidth]{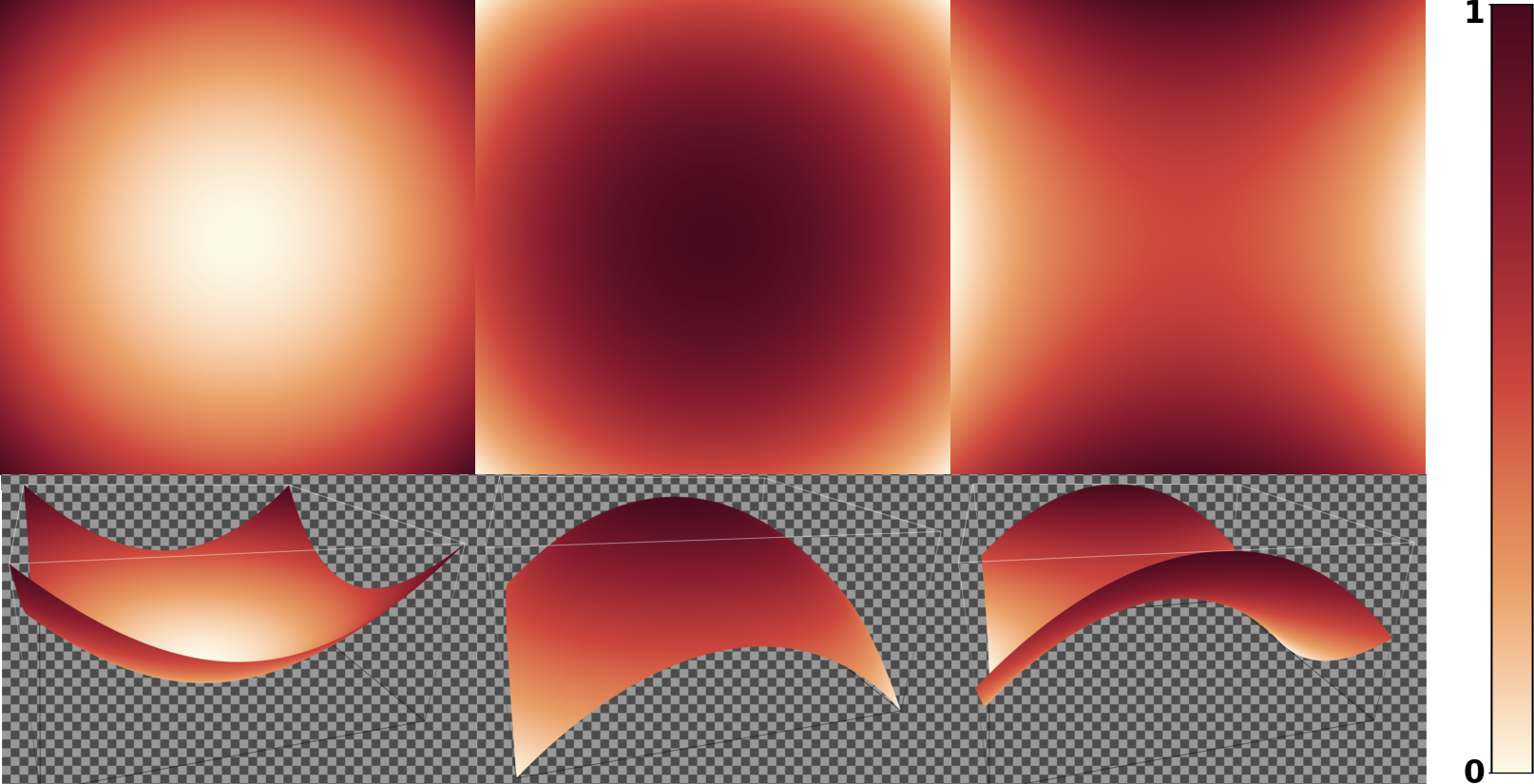}
  	\caption{ \label{fig:minMaxSaddleExample} This figure shows a 2D color mapping representation (top) and a 3D height-map (bottom) of the 2d scalar fields created with the test function yielding a minimum with $o=1$ and $p=1$ (left), a maximum with $o=-1$ and $p=-1$ (middle), and a saddle with $o=-1$ and $p=1$ (right).}
  \end{figure}
  
  \begin{figure}[t] 
  	\centering
  	\includegraphics[width=0.8\linewidth]{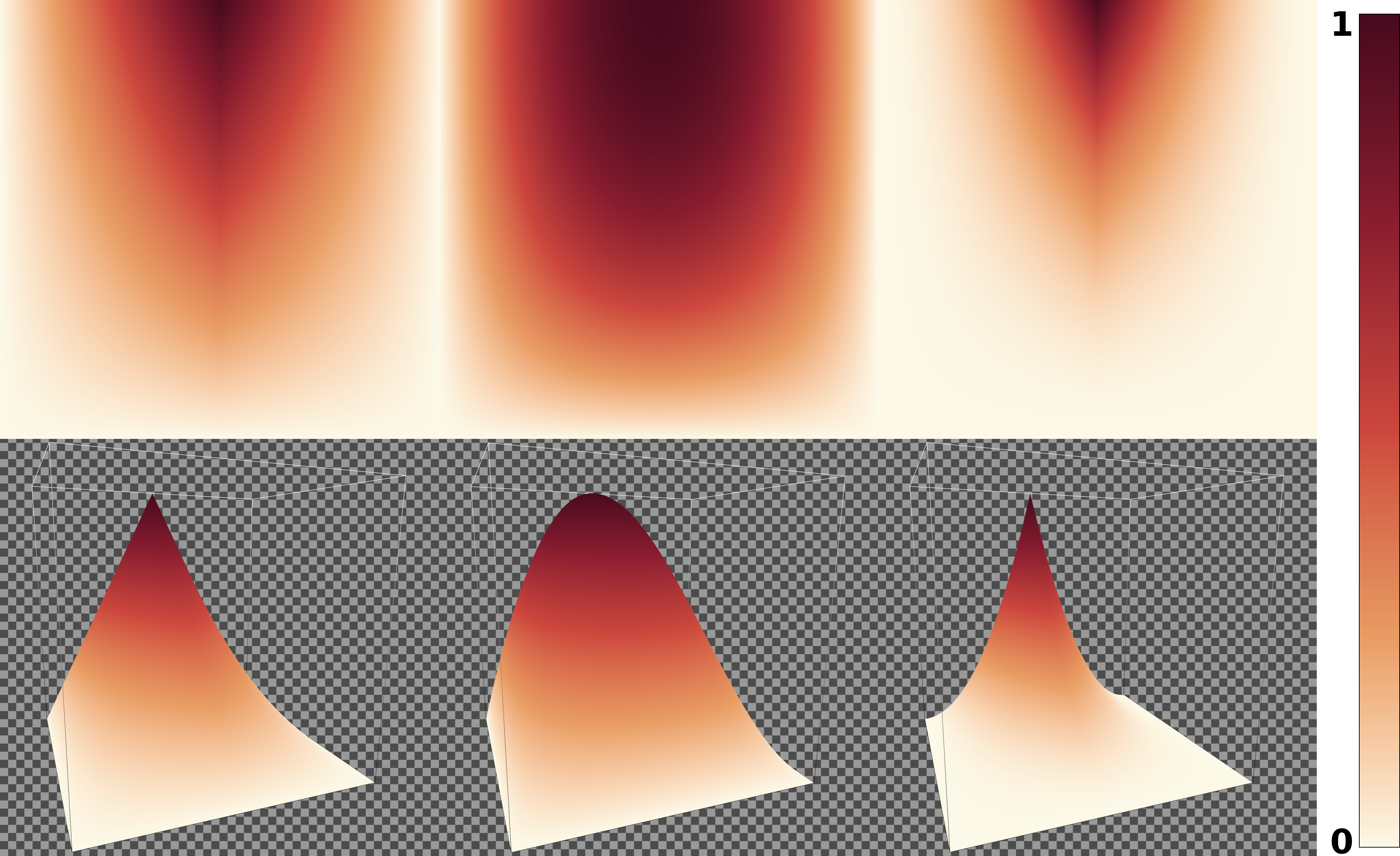}
  	\caption{ \label{fig:ridgeExample} Three ridge/valley-line tests (columns), with $r=0$, $R=1.0$. The ridge/valley-line is always centrally at $x=0$.
  	\textbf{Top Row}: Color mapping with $T_x=T_y=linear$, $T_x=T_y=concave$, and $T_x=T_y=convex$, and $b=2$ in the latter two cases.
  	\textbf{Bottom Row}: 3D height-map versions of the same tests.
  	}
  \end{figure}

\input{4_1_0_Local_Attributes.tex}

\input{4_2_0_Global_Attributes.tex}

\input{4_3_0_RealWorldData.tex}

%% file: 4_1_0_Local_Attributes.tex
\subsection{Local Tests}
Our basic design principle behind the local tests is classical calculus.
Local means that these test functions help to check the appearance of local properties of a scalar function after mapping it to color with the selected color map.
The main idea is to use typical local approximations like low order Taylor series expansions to create the test functions.
We use step functions to show the effect of discontinuities, and provide functions with different gradients, various local extrema, saddles, ridges, and valley lines.
This corresponds to ideas in the literature, as shown in \autoref{section:relatedwork}, e.g., works by Mittelst{\"a}dt~\cite{mittelstaedt2014methods,mittelstaedt2015colorcat} or Ware~\cite{ware1988color}.
We also use elements of Fourier calculus by providing functions to test the effect of different frequencies.
The final test looks at the colormap's potential to visually reveal small differences within the data range, which might be an important colormap design goal. %
  

 \input{4_1_1_Testing_Functions_NeighbourhoodVariations.tex}

 \input{4_1_2_Testing_Functions_GradientVariations.tex}
 \input{4_1_3_Testing_Functions_MinMaxSaddleVariations.tex}

 \input{4_1_4_Testing_Functions_Ridge_and_Valley_Lines.tex}

 \input{4_1_5_Testing_Functions_FrequencyVariations.tex}

 \input{4_1_6_Testing_Functions_TresholdVariations.tex}

%% file: 4_1_1_Testing_Functions_NeighbourhoodVariations.tex
%

\subsubsection{Step Functions} \label{subsubsection:neighbourhoodVariations}
Some popular test images in the literature use steps between adjacent pixels~\cite{mittelstaedt2014methods, mittelstaedt2015colorcat, Wang:2008}.
In terms of calculus, this means to use a function with discontinuities.
Ideally, the function should have different step heights starting from different levels.
For this purpose, we define a set $A ={a_0, ... , a_{n-1}}$ of increasing test values $a_i<a_{i+1}$.
The function is split into a rectangular grid with constant values.
Uneven columns contain in ascending sequence one value of $A$, while even ones contain $A$ in increasing order.
This means that some steps appear multiple times, but the function is rather simple to remember; see \autoref{fig:stepExample}.
Formally, the function $f_{Step}: [0,2n) \times [0,n) \to \mathbb R$  is given by

\begin{equation}
    f_{Step}(x,y)=
    \begin{cases}
        a_{\lfloor x \rfloor/2} & \textbf{if } \lfloor x \rfloor \mod 2 = 0\\
        a_{\lfloor y \rfloor} & \textbf{otherwise}
    \end{cases}
\end{equation}

\noindent where $\lfloor x \rfloor$ denotes the largest integer smaller or equal to $x$.
An observer can quickly identify the column, including one concrete test-value, and compare it with all values using the adjacent column.

%% file: 4_1_2_Testing_Functions_GradientVariations.tex
%

\subsubsection{Gradient Variation} \label{subsubsection:gradientVariations}

In the literature, the most common analytic example features gradient variation~\cite{borland2007rainbow, moreland2009diverging, mittelstaedt2014methods, mittelstaedt2015colorcat, bernard2015survey}.
In terms of calculus, at first glance, this may suggest using a first-order Taylor approximation.
However, the work by Ware~\cite{ware1988color} indicates that one should also test concave or convex properties, which means to use polynomials of somewhat higher-order, e.g., quadratic polynomials.

For our \texttt{Gradient Variation} test, we defined a test function $f_{Grad}(x,y):[0,1] \times [0,1] \to \mathbb{R}$ to examine different gradients including a concave or convex pattern.
The two options $T_x, T_y$ determine if the behavior is convex or concave along the $x$-axis or the $y$-axis.

Basically, along each horizontal line, we start at $r$ and interpolate to some value $g(y) \in \mathbb{R}$.
The function $g$ starts with $g(0)=r$ and ends with $g(1)=R$.
It may also use a linear, convex, or concave interpolation along $y$.
Its definition is

\begin{equation}
\label{equ:fct_Mapostrophe}
    g(y)=
    \begin{cases}
        (R-r) (1-(1-y)^{b}) + r,& \textbf{if } T_y = concave\\
        (R-r) y^b           + r,& \textbf{if } T_y = convex
    \end{cases}
\end{equation}

\noindent If $b$ equals one, then both cases describe a linear gradient along $y$, which we also denote as $T_y = linear$.
For $b \geq 2$, we get a concave shape with a decreasing gradient in the case $T_y = concave$ and a convex shape with increasing gradient for $T_y = convex$.

The actual test function is now defined in a similar manner by
\begin{equation}
\label{equ:gradient}
f_{Grad}(x,y)=
    \begin{cases}
        (g(y)-r) (1-(1-x)^{b}) + r,& \textbf{if } T_x = concave\\
        (g(y)-r) x^b + r,          & \textbf{if } T_x = convex
    \end{cases}
\end{equation}

\noindent Again, $b=1$ means a linear gradient along $x$ in both cases, and we can write $T_x=linear$ in this case.
$T_x=concave$ means a  convex shape created by a decreasing gradient, while $T_x=concave$ results in an increasing gradient and a concave shape.
With $R>r$, we get an increasing function and with $R<r$ a decreasing one.
\autoref{fig:gradientExample} shows 2D and 3D visualizations of examples for the function $f_{Grad}(x, y)$ for the three cases: $T_y=linear$, $concave$, and $convex$.
  
  \begin{figure}[t] 
  	\centering
  	\includegraphics[width=0.8\linewidth]{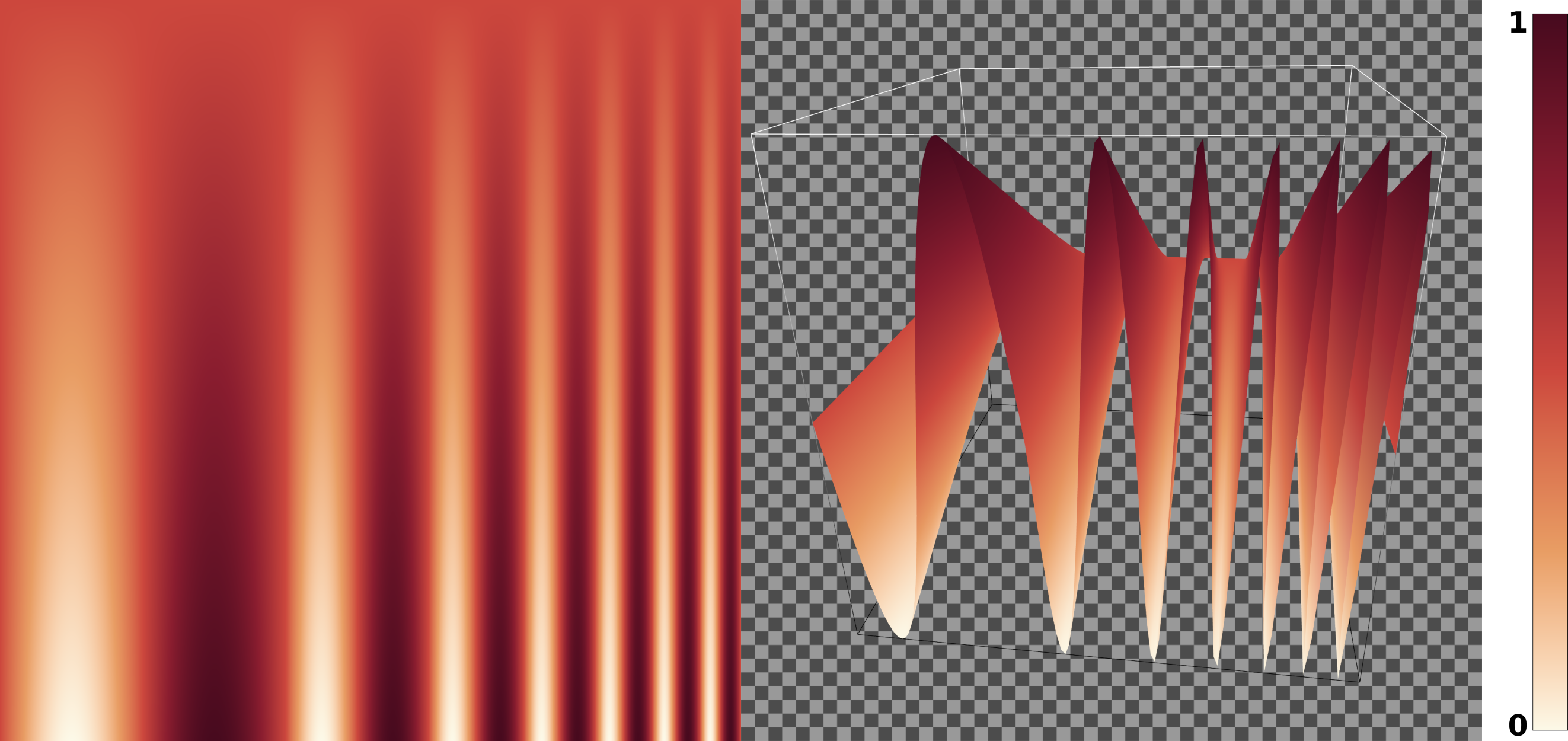}
  	\caption{ \label{fig:frequencyExample} 
  	    These two pictures show an example for the \texttt{Frequency Variation} test.
  	    Left: 2D color mapping of the test field.
  	    Right: 3D height-map visualization. 
  	    }
  \end{figure}
  
  \begin{figure}[t] 
  	\centering
  	\includegraphics[width=0.8\linewidth]{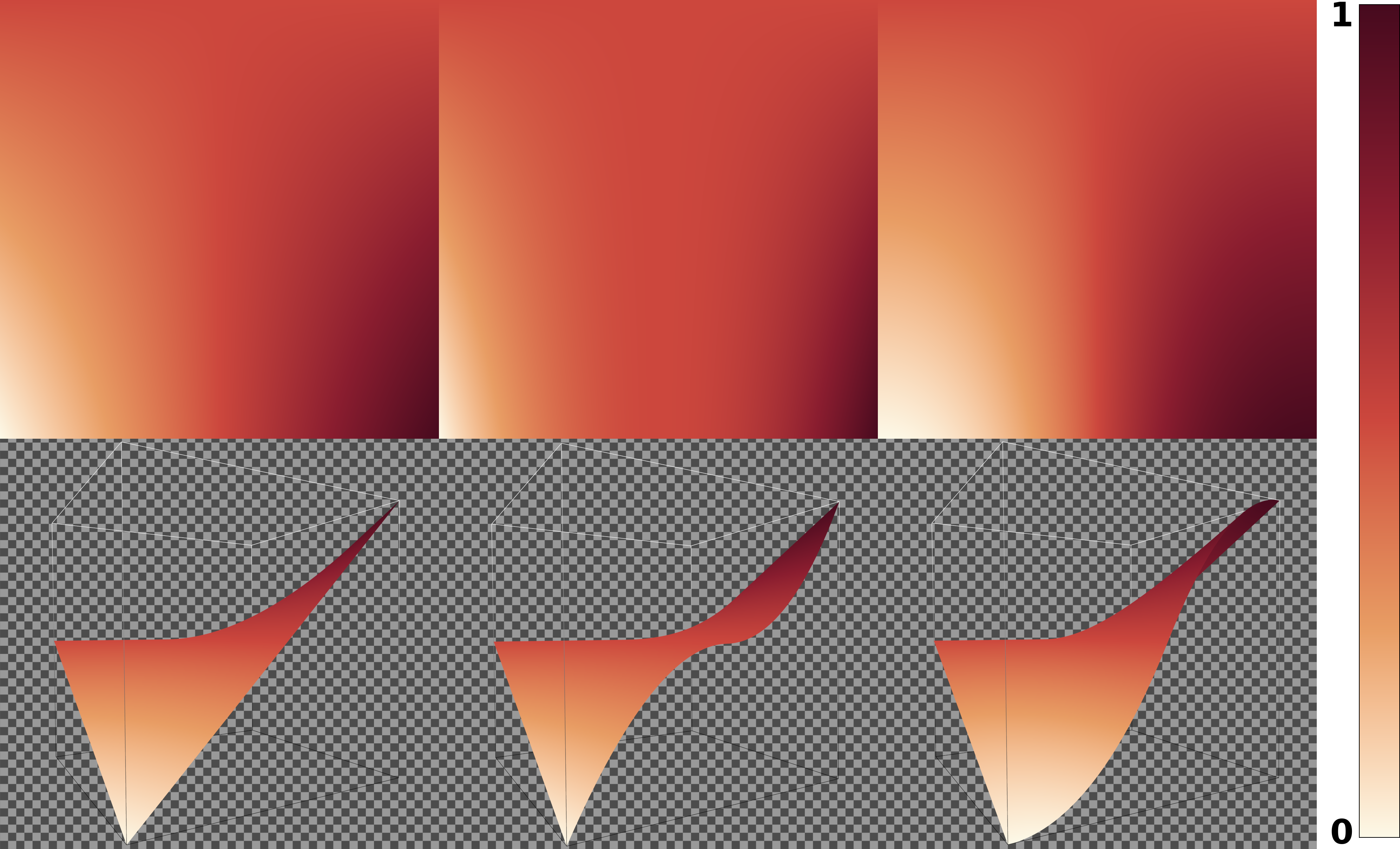}
  	\caption{ \label{fig:tresholdExample} This figure shows an example test image for each of the three \texttt{Threshold Variation} test types. From left to right you can see the linear, the flat, and the steep type. The top row shows the color mapping images and in the bottom row the corresponding test images as height-maps.}
  \end{figure}
  
  \begin{figure}[t] 
  	\centering
  	\includegraphics[width=\linewidth]{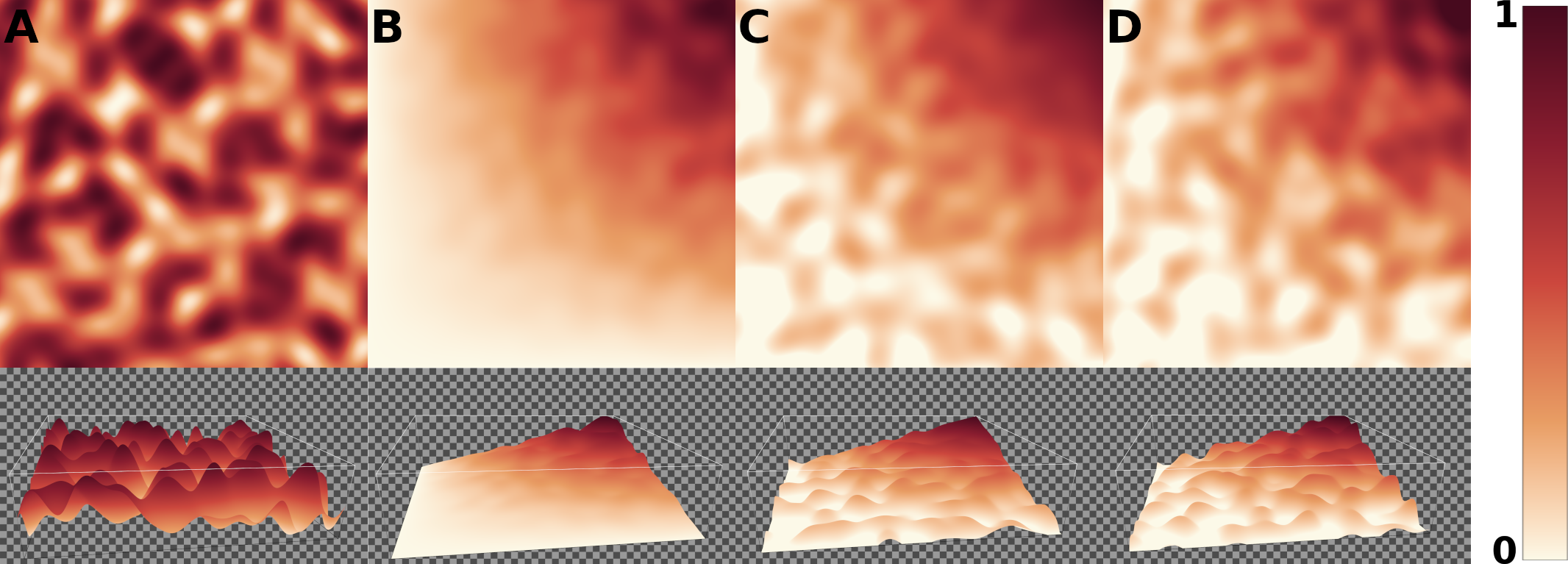}
  	\caption{ \label{fig:globalTopologyExample} This figure shows four color mappings of global topological structures with a gradient test function as foundation. The according 3D versions are shown at the bottom. The image \textbf{A} shows the pure noise using the $replacement$ option. The following images show the options $max-scaled$ (\textbf{B}), $min-scaled$ (\textbf{C}) and $range-scaled$ (\textbf{D}) with activated clipping and a limited value adjustment of 0.25. While \textbf{B} focused the noise to higher foundation values, \textbf{C} does the same with lower ones. In contrast \textbf{D} allows globally adjustments of the foundation.
  	}
  \end{figure}
  
  \begin{figure}[t] 
  	\centering
  	\includegraphics[width=0.85\linewidth]{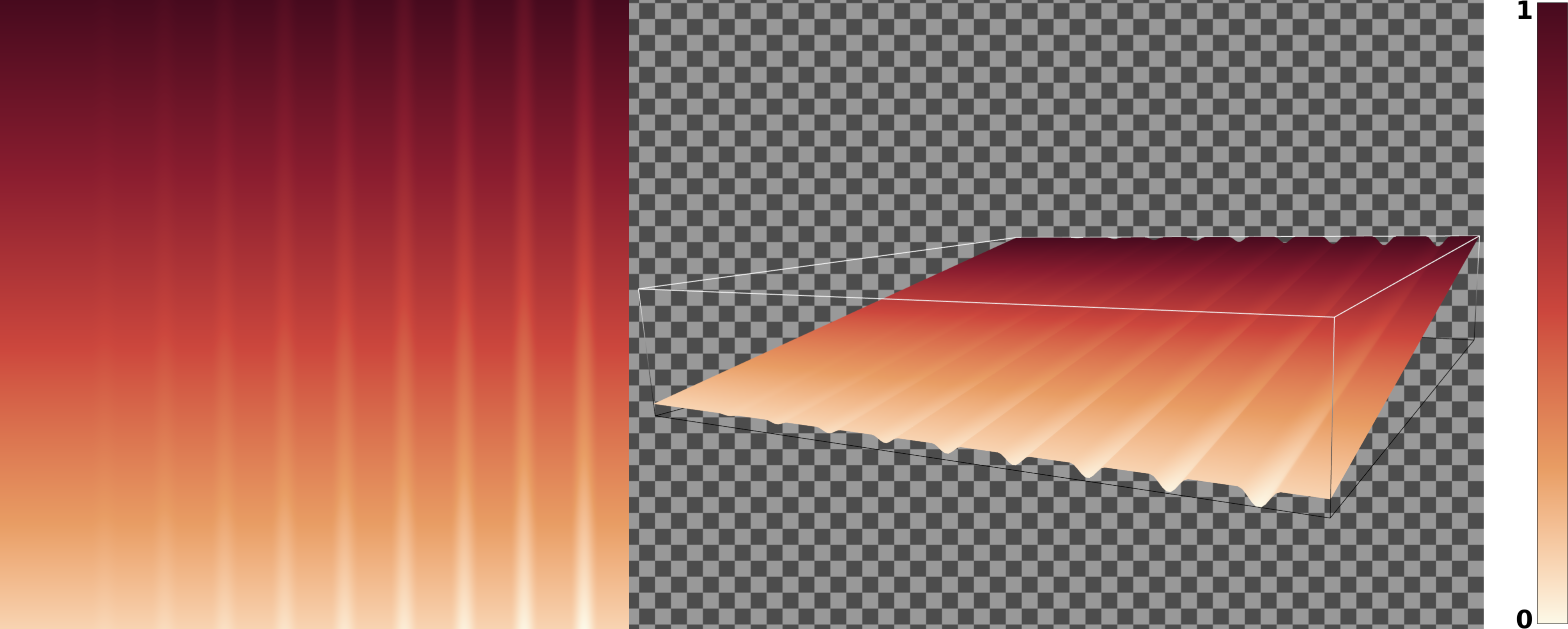}
  	\caption{\label{fig:littleBitExample}
  The two pictures show the \texttt{Little Bit Variation} test ($m=0.1$, $M=1$, $g_m=0.0001$, $g_M=0.1$), which can used to test the impression of small value variations by the visibility of vertical grooves. 
  The left image shows that the first groove with an amplitude of 0.0001 is not perceptible for all values between 0.1 and 1.0. 
  The right image shows a 3D height-map visualization of the same test.
  	}
  \end{figure}

%% file: 4_1_3_Testing_Functions_MinMaxSaddleVariations.tex
%

\subsubsection{Local Topology: Minimum, Maximum, and Saddle} \label{subsubsection:minMaxSaddle}

Besides discontinuities and gradients, local extrema and saddles are the next types of structures in scalar fields, according to calculus.
This has also been noted by Ware~\cite{ware1988color} and is the fundamental insight behind topological visualization methods~\cite{Heine:2016}.
As local extrema (of some significance or persistence) and saddles are of interest for most scalar field analysis tasks, we thus include them in our test suite.
For the calculation of user-definable minima, maxima, and saddles, we are using \autoref{equ:minMaxSaddle}.
  
\begin{equation}
  \label{equ:minMaxSaddle}
  f_{MMS}(x,y) = o  x^2 + p y^2 + m
\end{equation}

\noindent It is based on the well-known fact that stable critical points can be described by quadratic functions.
We put the extrema or saddle at zero.
The user can create a maximum with $o<0 \land p<0$, a minimum with $o>0 \land p>0$, and a saddle with $o>0 \land p<0 \lor o<0 \land p>0$.
The starting value of the structure is given by $m \in \mathbb{R}$.
\autoref{fig:minMaxSaddleExample} shows an example of this test function with visualizations of minima, maxima, and saddle points.

%% file: 4_1_4_Testing_Functions_Ridge_and_Valley_Lines.tex
%

\subsubsection{Local Topology: Ridge and Valley Lines} \label{subsubsection:ridgeValleyLines}

Besides local extrema and saddle points, ridge and valley lines are further relevant topological shape descriptors.
Again, this has been noted by Ware~\cite{ware1988color} with respect to color mapping.
Also, the relevance of ridges and valley lines is well established in feature-based flow visualization~\cite{Heine:2016}.
To test the suitability of colormaps for scalar fields that include such lines, we use a function $f_{RV}:[-1,1]\times[0,1] \rightarrow \mathbb{R}$.
The location of the ridge/valley-line is always at $x=0$ as a vertical line.
Its shape is determined by the function $g$ that we introduced in \autoref{subsubsection:gradientVariations}, so it may be linear, convex, or concave according to the exponent $b \in \mathbb{N}$ and the shape descriptor $T_y$.
For the slope in $x$-direction, we basically use the absolute value with exponent $b$, i.e., $|x|^b$ on the interval $[-1,1]$.
This creates a concave shape.
For convex shapes, we use the similar function $1-(1-|x|)^b$.
Both functions are adjusted to interpolate between $r$ at $-1$ and $1$ and $g(y)$ at $0$.
This is quite similar to the definition of $g$.
We introduce the type parameter $T_x$ and set it to "convex" or "concave" and arrive at the definition:

\begin{equation}
\label{equ:ridgeValley}
f_{RV}(x,y)=
    \begin{cases}
        (r-g(y)) |x|^b         + g(y) &  \textbf{if } T_x = concave \\
        (r-g(y)) (1-(1-|x|)^b) + g(y) &  \textbf{if } T_x = convex
    \end{cases}
\end{equation}

As in the gradient variation case, $b=1$ leads to the same linear function in both cases, which we also denote as "linear".
For $R>r$, we get a ridge-line. For $R<r$, we get a valley-line.
\autoref{fig:ridgeExample} shows an example test for ridge-lines using the different $T_x$ and $T_y$ types.

%% file: 4_1_5_Testing_Functions_FrequencyVariations.tex
%

\subsubsection{Frequency Variation} \label{subsubsection:frequencyVariations}

Another common test for analyzing colormaps involves variations of spatial frequencies~\cite{rogowitz1996not, borland2007rainbow, moreland2009diverging}.
To check the behavior for wave-like functions, the \texttt{Frequency Variation Test} uses an increasing frequency in the $x$-direction and a decreasing amplitude in the $y$-direction.
In $x$-direction, we use an additional parameter $D$ defining the number of frequency increases.
Basically, we start with a single sine wave with frequency $1$.
This takes one unit of space in $x$-direction.
Then we add a single sine wave of frequency $2$ using a half unit of space.
Then, we add a single sine wave of frequency $3$ using one third unit of space.
We continue in this way until we have $D+1$ waves of increasing frequency.
The $x$-range will obviously depend on the elements in the, so we denote their elements as

\begin{equation}
    x_0 = 0 \quad x_j = \sum_{k=1}^{j} \frac{1}{k}
\end{equation}  

\noindent The wave will swing around the median value $u$ with an amplitude $W$.
Now, we can define the test function by
  \begin{eqnarray}
    f_{Freq} : [x_0, x_{D+1}] \times [0,1] \to \mathbb{R} \\
    f_{Freq}(x,y) = 
      W (1-y) \sin(2 \pi j (x-x_{j})) + u, & x_{j-1} \le x \le x_j \nonumber\\
    j = 1,\ldots,D+1 \nonumber
  \end{eqnarray}
\noindent \autoref{fig:frequencyExample} shows an example of the \texttt{Frequency Variation} test with six different frequencies.
It should be noted that the image resolution is critical with regard to aliasing effects in case of high frequencies, i.e., when $D$ is large. 
%

%% file: 4_1_6_Testing_Functions_TresholdVariations.tex
%

\subsubsection{Threshold Variation} \label{subsubsection:tresholdVariations}

In many scientific disciplines, natural thresholds, such as the freezing point of water at $0^{\circ}${C}, and the data distribution close to them are of significant importance for visual analysis.
This is also a major reason for the high relevance of isolines and isosurfaces in scientific visualization.
  
  \begin{figure}[t] 
  	\centering
  	\includegraphics[width=\linewidth]{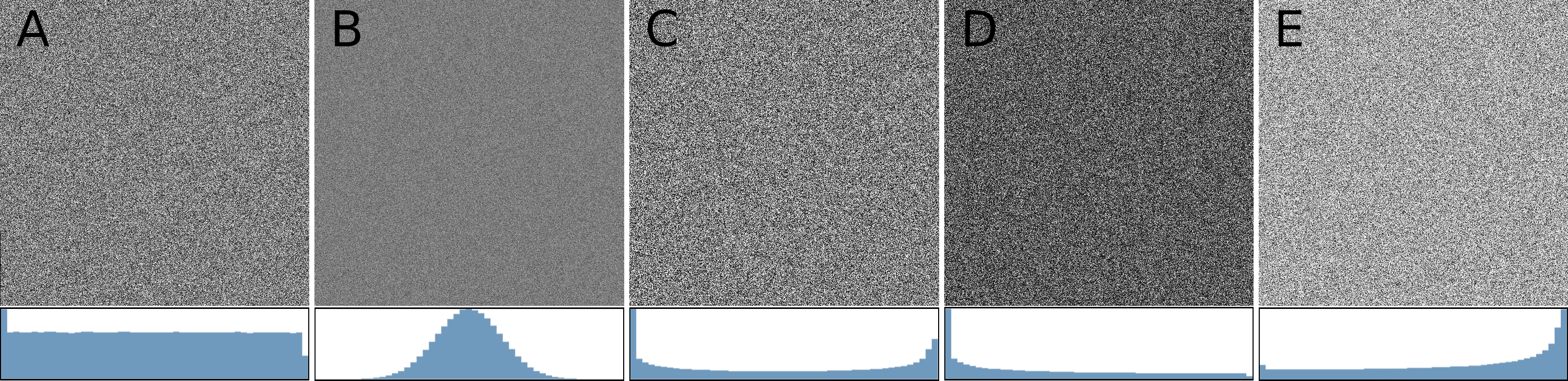}
  	\caption{\label{fig:noiseCollection}
  	    These images show the different distributions for the signal noise option. 
  	    To illustrate the noise, the pictures were created with the
  	      $replacement$ option and a noise proportion of 100\%. 
  	    You can see the following distributions:
          \textbf{1}=uniform,
          \textbf{2}=normal,
          \textbf{3}=beta,
          \textbf{4}=beta-left,
          \textbf{5}=beta-right.
        Below each picture is a histogram.
  	}
  \end{figure}
  
  \begin{figure}[t] 
  	\centering
  	\includegraphics[width=0.99\linewidth]{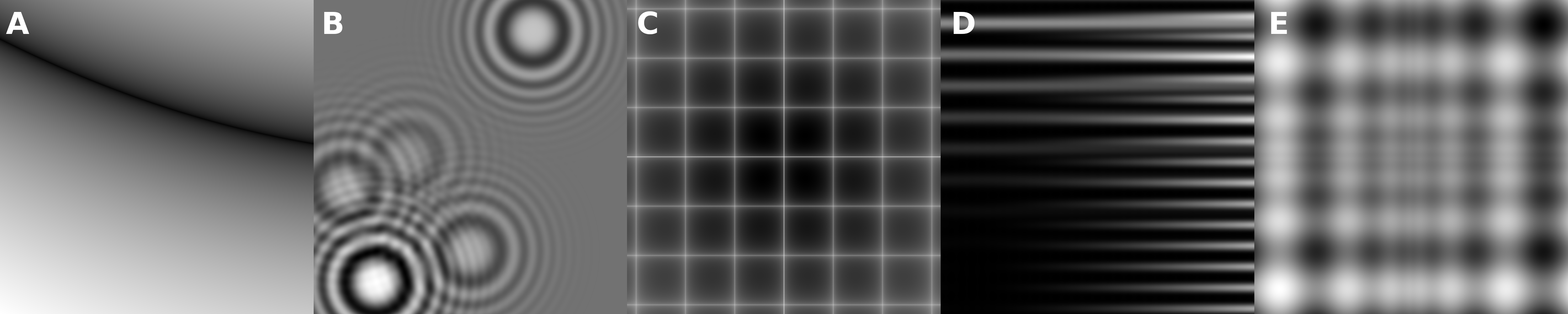}
  	\caption{\label{fig:collectionExample}
  	    This figure shows some examples for test functions used in the optimization computer science and also is of interest for our test function collection. 
  	    From left to right you can see the functions: 
  	    \textbf{A}:Bukin Function Number 6;
        \textbf{B}:Langermann Function;
        \textbf{C}:Cross-in-Tray Function;
        \textbf{D}:Levy Function Number 13;
        \textbf{E}:Schwefel Function. \cite{testfunctions::Molga::2005, testfunctions::Jamil::2013}
      	}
  \end{figure}
  
  \begin{figure}[t] 
  	\centering
  	\includegraphics[width=0.95\linewidth]{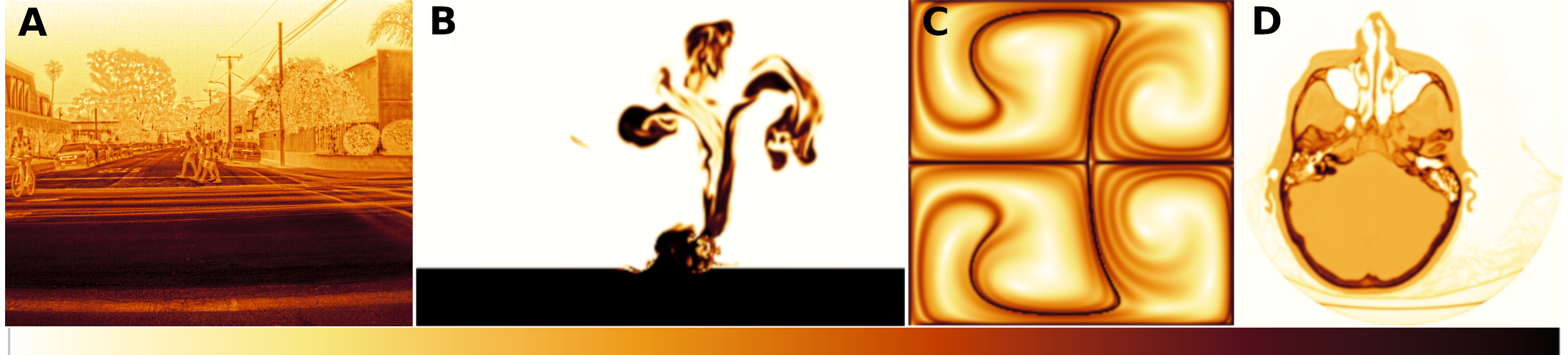}
  	\caption{\label{fig:realWorldExamples}
    This figure shows different real-world data sets we collected for our test suite. You can see the following data sets:
        \textbf{A}:A thermal data set for algorithm training (source: \url{https://www.flir.co.uk/oem/adas/adas-dataset-form/});
        \textbf{B}:Simulation data of asteroid impacts in deep ocean water (source: \url{https://sciviscontest2018.org/});
        \textbf{C}:FTLE technique from the field of flow visualization \cite{shadden2005definition};
        \textbf{D}:Medical computer tomography scan of a head (source: \url{https://graphics.stanford.edu/data/voldata/}).
  	}
  \end{figure}

 \begin{figure}[t] 
  \centering
    \includegraphics[width=0.95\linewidth]{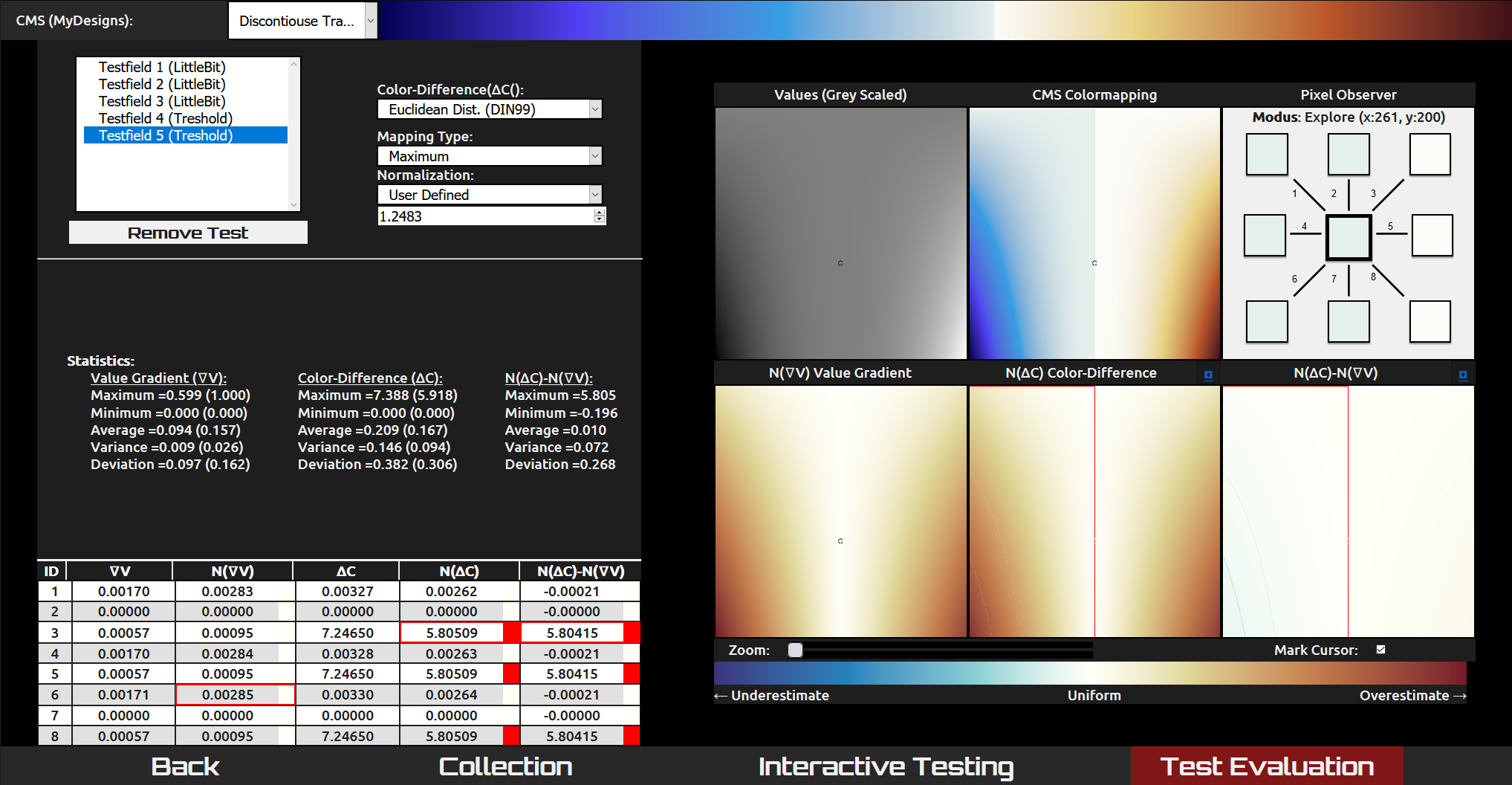}
  	\caption{\label{fig:evaluation_Screenshot}
  	  This figure shows a screenshot of the test evaluation in the 
  	    CCC-Tool. 
  	  We present statistics of the \texttt{Value Difference Field}, 
  	    \texttt{Color Difference Field}, and \texttt{Subtraction Field}. 
  	  On the right side are two visualizations of the test-function. 
  	  One color mapping is done with a grey-scaled colormap and the 
  	    other one with the selected colormap for the analysis. 
  	  Below are three color mapping images of the \texttt{Value Difference
  	    Field}, \texttt{Color Difference Field}, and \texttt{Subtraction Field}.
  	  All five images are zoomable and change interactively. 
  	  As a sixth part, the pixel observer shows in combination with the 
  	    table the pixel-neighborhood-information of the three fields.
  	}
 \end{figure}
  
It is possible to integrate isolines into a colormap by creating a discontinuous transition point in the colormap~\cite{nardini2019making}.
With the \texttt{Threshold Variation} test, we created a function for testing a specific user-defined threshold $t$.
The test function $f_T$ spans the domain $[-1,1] \times [-1,1]$.
The isoline with the value $t$ is a vertical line in the middle, i.e., $x=0$.
On the left, the values are lower, and on the right, they are higher.
We use the two parameters $m$, and $M$, with $m<t<M$.
To increase visual information in the test image, the functions $f_{M}, f_m : [-1,1] \to \mathbb{R}$ change the minimum and maximum value linearly depending on $y$:
\begin{equation}
\label{equ:fct_MapostropheTresh}
    f_M(y)= \frac{M+t}{2} - \frac{M-t}{2} y
\end{equation}
\begin{equation}
\label{equ:fct_MapostropheTresh_m}
    f_m(y)= \frac{t+m}{2} + \frac{t-m}{2} y
\end{equation}
\noindent We can now define our test function $f_{T}:[-1,1] \times [-1,1] \to \mathbb{R}$.
Along each horizontal line, i.e., for fixed $y$, the value changes from $f_m(y)$ at $-1$ to $t$ at $0$, and finally reaches $f_M(y)$ at $1$.
We use three cases that we call $T=linear$, $T=flat$, and $T=steep$. 
The type describes the gradient behavior around the threshold. 
As in the cases for the gradient variation and ridge/valley lines, we use an exponent $b \in \mathbb{N}$ as final parameter.
$b=1$ describes the linear case, while higher exponents create a rather flat or steep gradient around the isoline.
The formal definition of our test function is

\begin{multline}
  \label{equ:treshold}
  f_{T}(x,y)= \\
    \begin{cases}
      (f_m(y)-t) |x|^b         + t & \textbf{if } (T=flat \land x \leq 0)\\
      (f_M(y)-t) |x|^b         + t & \textbf{if } (T=flat \land x > 0)\\
      (f_m(y)-t) (1-(1-|x|)^b) + t & \textbf{if } (T=steep \land x \leq 0) \\
      (f_M(y)-t) (1-(1-|x|)^b) + t & \textbf{if } (T=steep \land x > 0)
    \end{cases}
\end{multline}

\noindent In \autoref{fig:tresholdExample}, a plot shows examples of all three types.

%% file: 4_2_0_Global_Attributes.tex
 \subsection{Global Tests}

In contrast to the local tests, the global tests look at more global properties of scalar functions and how well the colormap presents them.
First, we look at global topological properties.
We use functions showing Perlin noise to create multiple local minima and maxima on different height levels and different spatial structures.
Details can be found in \autoref{subsubsection:globalTopologicalStructures}.
Second, it is a challenge for any colormap to deal with a large span of the overall values while small, but relevant, local value variations are also present.
Any nearly linear colormap will completely overlook these so-called little bit variations.
As test functions, we use linear functions with varying gradient and height as background with small grooves to include little bit variations.
The definition is given in \autoref{subsubsection:littleBitVariations}.
Third, real-world data, especially images created by measurements, contain noise of various types and intensity, i.e., signal-to-noise ratio.
We use functions from the local test suite and add uniform or Gaussian distributed noise of different signal-to-noise ratios.
We describe the details in \autoref{subsubsection:signalNoiseVariations}.
Finally, we add a collection of test functions from other computer science disciplines to allow for tests using these functions.

\input{4_2_1_Global_Topological_Structures.tex}

\input{4_2_2_Testing_Functions_LittleBitVariations.tex}

\input{4_2_3_Testing_Functions_SignalNoiseVariations.tex}

\input{4_2_4_Testing_Functions_Collection.tex}

%% file: 4_2_1_Global_Topological_Structures.tex
%

\subsubsection{Global Topological Structures} \label{subsubsection:globalTopologicalStructures}

As noted above, other authors indicated the relevance of critical points for testing colormaps before.
In contrast to the local topology in \autoref{subsubsection:minMaxSaddle}, we use a larger number of critical points in the following test.
For the creation of global topological structures, we take the 2D version of the improved noise algorithm introduced by Perlin \cite{Perlin_1985, Perlin_2002}, which is often used for the creation of procedural textures or for terrain generation in computer games.
The idea of this test function is to use some other test function from \autoref{subsubsection:gradientVariations} to \autoref{subsubsection:littleBitVariations} as a background and combine this field with noise according to Perlin's work.
These distorted gradients and shapes are in analogy with colormap testing functions specifically used to determine the discriminative power of subregions of colormaps~\cite{rogowitz1999trajectories, kalvin2000building, ware2017evaluating, ware2017uniformity, moreland2009diverging}.
  
To create the critical points, we use the noise function $f_{Noise}(x,y) \in [-n,n]$, $n>0 \land n\leq1$ and distinguish four options (see \autoref{fig:globalTopologyExample}).
For the options $min-$,$max-$, and $range-scaled$ the selected test-function $f_{test}$ affect the result. Whereby the influence of the first two options depend on the closeness of the local value $f_{Noise}(x,y)$ to the $m$, or rather $M$.
This procedure creates noise that is focused on small/high values.
At the $range-scaled$ option, the adjustment of the local value is limited by the test-function range from $m$ to $M$.
Furthermore, an optional clipping method for these three options prevent values out of $[m,M]$.
Fourth, we offer $replacement$  as a final option, where users can set a custom noise-range $N=[n_m,n_M]$ with $f_{Noise}(x,y) \in N$.
With this option, the entries of the test-function will be replaced by the noise value.

\begin{equation}
\label{equ:noiseOptions}
f_{test}(x,y)=
    \begin{cases}
        f_{test}(x,y)+f_{Noise}(x,y)*\frac{f_{test}(x,y)-m}{M-m} & \textbf{if } max-scaled \\
        f_{test}(x,y)+f_{Noise}(x,y)*\frac{M-f_{test}(x,y)}{M-m} & \textbf{if } min-scaled \\
        f_{test}(x,y)+f_{Noise}(x,y)*(M-m) &  \textbf{if } range-scaled \\
        f_{Noise}(x,y) &  \textbf{if } replacement
    \end{cases}
\end{equation}

%% file: 4_2_2_Testing_Functions_LittleBitVariations.tex
%

\subsubsection{Little Bit Variation} \label{subsubsection:littleBitVariations}

The teaser \autoref{fig:camelExample} demonstrates that standard colormaps may easily lead to overlooked small value variations.
For such cases, i.e., if small variations in the scalar field (within a small sub-range of the full data range) carry valuable information for interpretation, we define a test on the potential of a given colormap to visually resolve small perturbations.
This is similar to distorted gradients, which appear quite frequently in the literature~\cite{rogowitz1999trajectories, kalvin2000building, ware2017evaluating, ware2017uniformity, moreland2009diverging}.
The \texttt{Little Bit Variation} test
 
\begin{equation}
    f_{LB}:[0,2n+1]\times[0,1] \rightarrow \mathbb{R}
\end{equation}

\noindent uses a background function and adds a function $f_{G}$ producing $n$ small grooves. 
The background function in this test is a linear gradient along the y-direction, which is defined by a user-specified value range $[m,M]$.
Along the $x$-direction, this function is modified by a function $f_{G}$ creating $2n+1$ alternate stripes of unchanged background and grooves, so we use

\begin{equation}
 \label{equ:littleBit}
    f_{LB}(x,y)=m+(M-m)y-f_{G}(x)
\end{equation}

\noindent The function $f_G$ produces sine-shaped grooves for odd $\left\lfloor x \right\rfloor$ and no changes for even $\left\lfloor x \right\rfloor$.
As $x$ runs from $0$ to $2n+1$, this creates exactly $n$ grooves.

\begin{equation}
\label{equ:littleBitGroove}
    f_{G}(x) =
    \begin{cases}
         0,                      & \textbf{  if } \lfloor x \rfloor \mod 2 =0\\
         - f_{A}(x) \sin( \pi (x - \lfloor x \rfloor)), &  \textbf{otherwise}\\
    \end{cases}
\end{equation}

\noindent As can be seen, the sine wave's amplitude is changed by a function $f_{A}(x)$ and create a test of different small value changes (groove depths).
The function $f_{A}(x)$ determine for each groove the depth by linear interpolation between user-defined minimum $g_m$ and maximum $g_M$.
In \autoref{fig:littleBitExample}, you can see an example for the \texttt{Little Bit Variations} test. 

\begin{equation}
\label{equ:littleBitAmplitude}
    f_{A}(x) = g_{m} + \frac{\lfloor x \rfloor - 1}{2*n-2} (g_{M}-g_{m})
\end{equation}

%% file: 4_2_3_Testing_Functions_SignalNoiseVariations.tex
%

\subsubsection{Signal-Noise Variation} \label{subsubsection:signalNoiseVariations}

In the signal and data processing, noise plays an important role.
It also affects the results of scientific visualizations.
Like the global topology test  (see~\autoref{subsubsection:globalTopologicalStructures}), our tool offers to add noise to each test function (\autoref{subsubsection:gradientVariations} - \autoref{subsubsection:littleBitVariations}).
The tool uses the standard random algorithm from JavaScript, which produces pseud- random numbers in the range $[0,1]$ with uniform distribution.
For the noise behavior, we offer the same noise behavior options from \autoref{subsubsection:globalTopologicalStructures}.
Independent from the selected option, the fraction of noisy pixels can be set.
This fraction describes how many randomly selected field values are affected by noise.
If the noise proportion is set to 100\%, the full test-function is affected by noise.
For more flexibility, we also offer a conversion from a uniform distribution to a normal or a beta distribution.
The conversion from uniform to the normal distribution is done with the Box-Muller transform~\cite{BoxMull58}.
With the normal distribution, the noise will be more focused on weaker changes around null for the $min/max-scaled$ and $range-scaled$ options.
For the $replacment$ option, the normal distribution causes a focus on values around the median of the defined range of noise values.
The approach from uniform to a beta-like distribution (with $alpha,beta=0.5$) is done with the equation $beta_{Random}=sin(r*\frac{\pi}{2})^2$, with $r$ being the result of the standard random generator.
Adding noise using a beta distribution with the $min/max-scaled$ or $range-scaled$ options will have a priority for values near the maximal change parameter $m$ and $-m$.
For the $replacement$ option, values near the minimum and maximum of the defined noise value range will be preferred.
We modified this conversion with a view to do this preference on only one side, thus for $m$ or $-m$ in the first case or for the maximum or minimum in the other case.
The modification is a mirror at the median random value to the left or right side of this median.
This allows us to create a beta-like distribution and also a left-oriented beta distribution and a right-oriented beta distribution.
\autoref{fig:noiseCollection} shows the different distribution options.

%% file: 4_2_4_Testing_Functions_Collection.tex
%

\subsubsection{Function Collection} \label{subsubsection:testcollection}

Many domains of computer science use test functions for the evaluation of algorithms.
There are several widespread well-known functions like \texttt{Mandelbrot Set} or \texttt{Marschner Lobb} and also functions like the \texttt{Six-Hump Camel Function} from the teaser, which are better known in optimization than computer science \cite{Mandelbrot::1980, testfunctions::MarschnerLobb::94, testfunctions::Jamil::2013}.
Such functions and their different attributes could also be an enrichment for evaluation in scientific visualization.
Therefore, we included a collection of such functions from the literature in our testing environment.
These functions stand beside our development of test functions, and provide further challenges for colormaps.
With this collection, we want to provide over time more and more such functions of interest.
In order to allow users to test their colormaps without changes, we allow the user to scale the values of these functions to the range of the colormap or a user-defined range.
\autoref{fig:collectionExample} shows some examples of functions used for optimization.
Obviously, they also have relevant properties for the evaluation of color mapping.
For example, the \texttt{Bukin Function} includes many small local minima along a valley-line. \cite{testfunctions::Jamil::2013}

%% file: 4_3_0_RealWorldData.tex
%

  \begin{figure}[t] 
  	\centering
  	\includegraphics[width=0.90\linewidth]{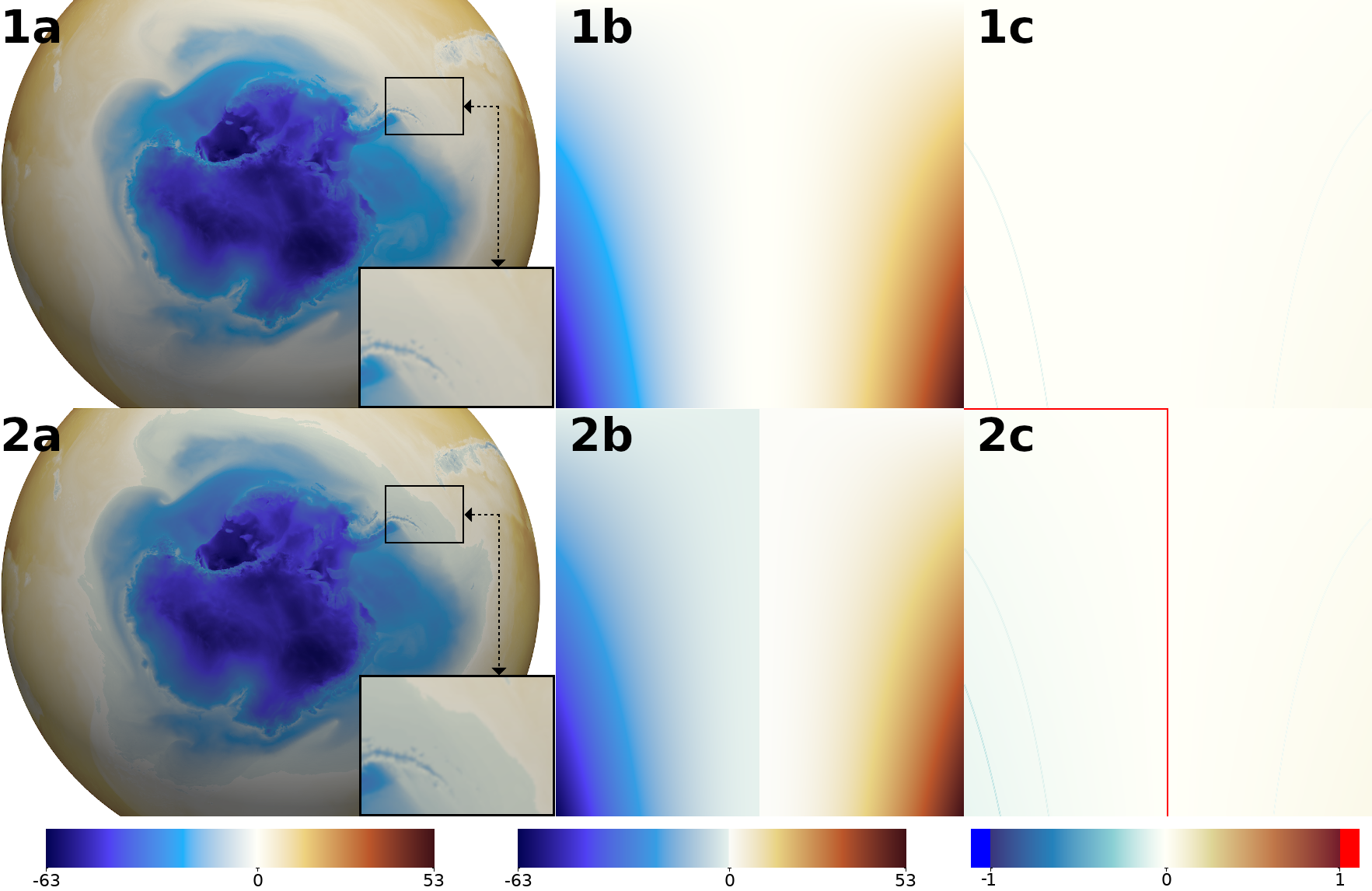}
  	\caption{\label{fig:application_Threshold}
  	    This figure shows application of the $Threshold$ test (\autoref{subsubsection:tresholdVariations}) to improve the distinguishability of temperature variations around the South Pole with focus on the freezing point (see close-ups in the lower right corner of \textbf{a}). \textbf{1}: Local uniform optimized cool-warm colormap. \textbf{2}: Modified colormap with a discontinuous transition point to improve the threshold. \textbf{a}: Visualization of the 2m-temperature of a high resolution simulation with the global atmosphere model ICON. \textbf{b}: Threshold test function with $m=-63$, $M=53$, and $t=0$. \textbf{c}: \texttt{Subtraction Field} of the evaluation method (\autoref{section:testevaluation}). 
  	}
  \end{figure}
  
  \begin{figure}[t] 
  	\centering
  	\includegraphics[width=0.90\linewidth]{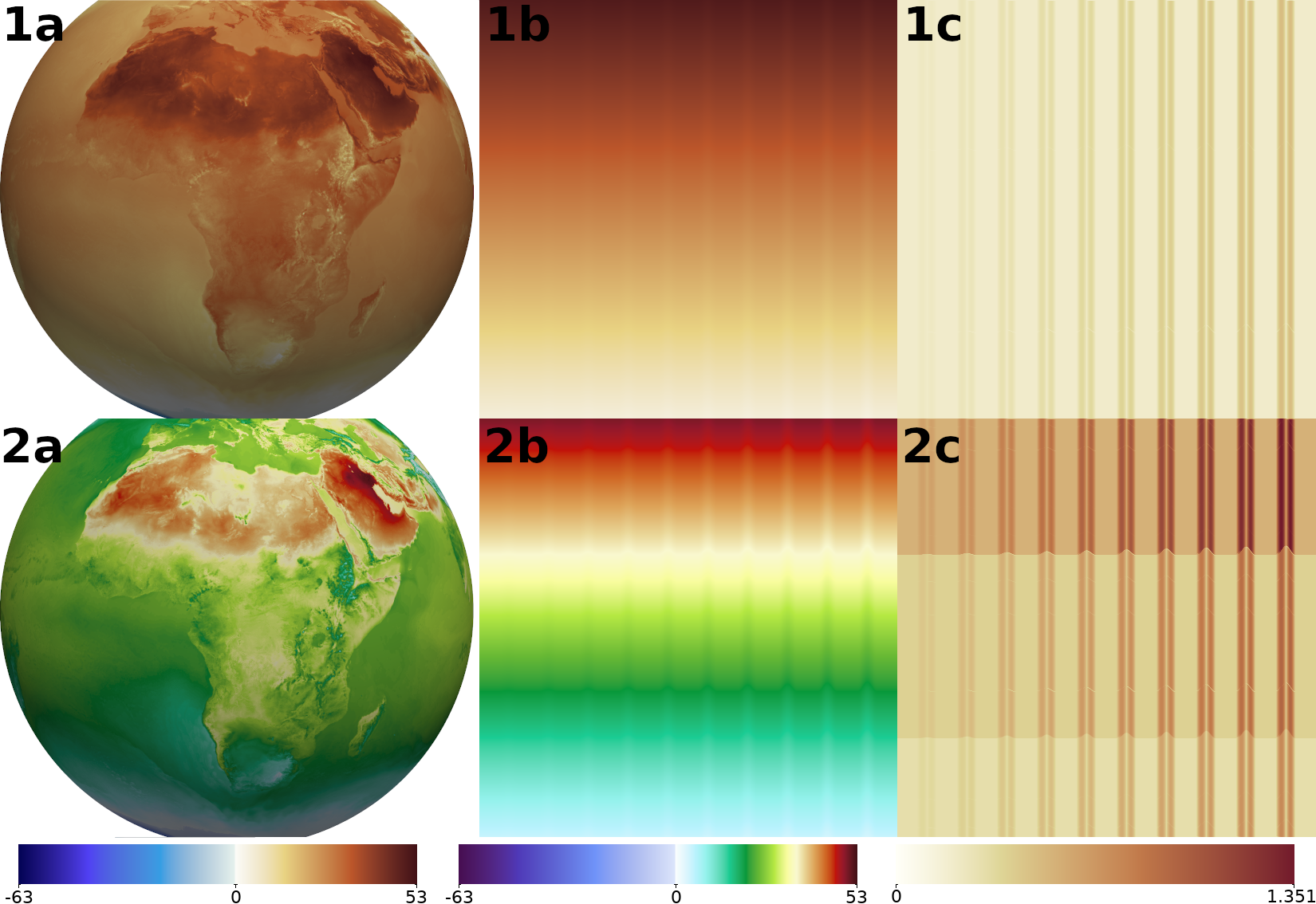}
  	\caption{\label{fig:application_LittleBit}
  	  \textbf{1}: Starting with the modified colormap of \autoref{fig:application_Threshold},  we used the \texttt{Little Bit} test (\autoref{subsubsection:littleBitVariations}) to increase the number of noticeable values of positives temperatures. The view in the \textbf{a}-panels is centered on Africa.  
  	  \textbf{2}: Modified colormap which uses more hue variations to improve the \texttt{Little Bit} results. 
  	  \textbf{a}: Visualization of the $2m$-temperature of a high resolution simulation with the global atmosphere model ICON. 
  	  \textbf{b}: \texttt{Little Bit} test function with $m=5^{\circ}${C}, 
  	    $M=53^{\circ}${C}, $g_m=0.1^{\circ}${C}, and $g_M=1.0^{\circ}${C}.
  	  \textbf{c}: \texttt{Color Difference Field} of the evaluation 
  	    method (\autoref{section:testevaluation}).
  	}
  \end{figure}

\subsection{Application-Specific Tests} \label{subsection:realWorldData}
In the two previous sections, we described several analytic test functions concerning specific challenges encountered in color mapping.
Additionally, we also introduced a collection of already existing test functions from other computer science domains.
Nevertheless, we think that the involvement of real-world data is indispensable for the completion of this test suite.
Real-world data originates from many different sources, is generated with various measurement techniques or simulation algorithms, and includes a myriad of attribute variations.
Most importantly, such data could potentially present several of the challenges described in the two previous sections at the same time. 
This kind of test cannot be easily replaced by our theory-based test functions completely.  
Therefore, we decided to include a set of application test data from different domains to cover a wide spectrum of realistic challenges.

Within one specific scientific domain, there is often a similarity between typical data sets; e.g., in medicine, data from the MRI (Magnetic Resonance Imaging) or the CT (Computer Tomography) is frequently used. 
Such data sets have similar attributes, and similar requirements have to be fulfilled by colormaps.
If we cover different typical data sets of different scientific disciplines, in the future, we can hopefully offer enough different real-world test cases so that most users will find a case that has some similarities with his data.
Like the test function collection from \autoref{subsubsection:testcollection}, this collection of real-world data will be extended over time.
At the current version, the tool offers medical-, flow-, and photograph-specific real-world data.

%% file: 5_0_TestEvaluation.tex
\section{Test Evaluation} \label{section:testevaluation}

Mostly there are good reasons to select specific colormaps or to design colormaps in a specific way.
Depending on the actually envisaged purpose of the colormap, a user decides on the number of keys; the hue, saturation and value of each key; the gradients in the mapping between the data range and the colormap; and so on.
Furthermore, \emph{de facto} standards and cognitive motivation may also influence the user's choice. 
Therefore, meaningful automated evaluation of continuous colormaps without knowledge of their intended use is rarely feasible.
Therefore, a general colormap score computed based on automatic tests and benchmarks might not be informative.

Instead, we propose to derive information based on aforementioned test functions that can be analyzed and rated by users themselves.
A user first chooses a test-function from \autoref{section:testingFunctions}.
For each grid point of the generated test field, we calculate the value differences to the neighboring grid points.
Depending on the location within the field, the number of neighbors varies between three and eight.
We normalize these value differences with the minimum and maximum value differences found and save them into a \texttt{Value Difference Field}.
We repeat this process also for the colors.
Here, we use some color difference norm (Lab, DIN99, DE94, or CIEDE2000) and save the normalized values into the \texttt{Color Difference Field}.

By subtracting these two fields from another, we get a \texttt{Subtraction Field}.
This field represents the local uniformity of the color mapping; when the local gradients found in the data are accordingly represented in the color mapped field, the difference between normalized data field and normalized color mapped field is zero for all pixels/locations. 
In the case of a non-linear color mapping, in contrast, the \texttt{Subtraction Field} will particularly highlight areas with strong non-linear mapping, which the user might have designed intentionally in order to increase the number of discriminable colors for a part of the data range.
The user can study the \texttt{Color Difference Field} as well as the \texttt{Subtraction Field} to analyze the color mapping of the test function.
%

Each of the three fields has three up to eight values for each pixel.
For the color mapping (\autoref{fig:evaluation_Screenshot}), the user can select maximum, average, or median.
Next to that, there are options to select a method for the calculation of the color difference.
The tool offers Euclidean distance for Lab and DIN99 space or the use of the DE94 or CIEDE2000 metrics in the Lab space. 

To compare the visualizations of \texttt{Color Difference Field} of different colormaps, we cannot use the normalization by minimum and maximum. The colors of such color mappings would relate to different color difference values and are not comparable.
Therefore we implemented two alternative options using fixed values for the minimum and maximum of the normalization to create comparable results.
The \texttt{Black-White} normalization use the greatest possible color difference between black and white as maximum and zero as minimum.
The \texttt{Custom} normalization uses a user-entered maximum, which is a necessity if the black-white difference is to big by contrast with the occurring color differences of the \texttt{Color Difference Fields}.
In \autoref{fig:application_Threshold}, we used this third option to get a comparable visualization for a colormap with a discontinuous transition point.

%% file: 6_0_applicationCase.tex
\section{Application Case} \label{section:application}

In this section, we show how the test suite could be utilized to evaluate the suitability of colormaps with respect to a given application problem.
For this example, we chose a data set from a simulation with a high-resolution global atmosphere model.
The data we use is one timestep of the temperature at a height of 2m simulated with the icosahedral ICON model at a global resolution of 5km~\cite{Stevens:2020}.
We remapped the data from the unstructured model grid to a regular grid with $4000 \times 2000$ grid points for easier use with different tools. 

On the global scale, the 2m-temperature is typically characterized by a wide range of values between less than $-80^{\circ}${C} and more than $50^{\circ}${C}.
For the selected time step, the simulated 2m-temperature varies between about $-63^{\circ}${C} and $52^{\circ}${C}.
Regionally, however, small temperature variations of the order of $0.1^{\circ}${C} might be critical for the analysis as, e.g., in the neighborhood of the freezing point at $0^{\circ}${C}.

Panel \textbf{1a} of \autoref{fig:application_Threshold} shows a visualization of the data using a spherical projection with a focus on the South Pole.
In contrast to mountainous regions, where the horizontal $2m$-temperature gradient is generally high, the gradient in flat areas such as oceanic regions is much smaller.
Here, the color differences are too small to depict local temperature variations, as for example, in regions with values close to $0^{\circ}${C} as shown in the close-up in the lower right corner of the image.

To test a given colormap for its discriminative power in the data range around the freezing point, we applied a threshold test with the options $Flat-Surrounding$, $m=-63$, $M=53$, and $t=0$.
First, we start with a local uniform cool-warm colormap (\textbf{1a} of \autoref{fig:application_Threshold}).
The related test function visualization \textbf{1b} demonstrates that it is impossible to differentiate between negative and positive values if the values are close to $0^{\circ}${C}.
The \textbf{1c} \texttt{Subtraction Field} method of the test evaluation part (\autoref{section:testevaluation}) yields a nearly white image, which reflects that the colormap uniformly represents the gradients produced by the test function.
To highlight the freezing point in the mapping, we introduce a non-linearity in the colormap, at $0^{\circ}${C}.
We use the twin key option of the CCC-Tool colormap specification (CMS)~\cite{nardini2019making}, which separates the color key at $0^{\circ}${C} into a left and right color key to create the discontinuous transition.
To improve the visual difference between both sides, we slightly lower the lightness value and increase the left color saturation to achieve light blue.
We kept white as the right-hand part of the color key.
Panels \textbf{2a} and \textbf{2b} of \autoref{fig:application_Threshold} illustrates that the introduced discontinuity in the colormap clearly separates the areas with negative and positive temperature values.
In comparison to \textbf{1c}, the \texttt{Subtraction Field} in \textbf{2c} shows with a vertical red line the spatial position of the discontinuous transition at $0^{\circ}${C}.
The according visualization of the temperature field of the modified colormap is shown in panel \textbf{2a}.

If we visualize the global 2m-temperature field using a linear colormap and look at the tropics or the mid-latitudes, we find that regional variations are also not very well resolved.
Using the same colormap, \autoref{fig:application_LittleBit}~\textbf{1a} shows a different view onto our planet, as \autoref{fig:application_Threshold}~\textbf{1a}.
The resolving power of the linear colormap is equally distributed over the full data range.
However, when we analyze the global temperature distribution, we find that more than half of the data range is used for the temperature variations far below $0^{\circ}${C} mostly in Antarctica, although this information is less important for most users of such a data set.
With respect to vegetation and agriculture, we may want to put more focus on regions with temperatures mostly above $0^{\circ}${C}.

Therefore we extended the path of the colormap through the color space to get more distinguishable colors for the positive data range.  
We used a \texttt{Little Bit} test to control improvements during this process.
Panel~\textbf{1a} of \autoref{fig:application_LittleBit} shows a visualization using the colormap with the discontinuous transition introduced above.
The corresponding \texttt{Little Bit} test is shown in panel~\textbf{1b}.
For the evaluation, we used the \texttt{Color Difference Field} (\autoref{section:testevaluation}).
Panel~\textbf{1c} shows how the small grooves in the linear gradient of the \texttt{Little Bit} test function (that are hardly noticeable in \textbf{1b}) become clearly visible in the color difference field.
From left to right, the regularly spaced perturbations in the field increase in magnitude, which is represented by a stripe pattern in panel~\textbf{1c} that increases in contrast from left to right.
The vertically constant color of the stripe pattern is a direct consequence of a linear colormap.

However, as we wanted to increase the discriminative power in the upper part of the colormap, we inserted additional color keys.
First, we moved the blue part of the colormap representing negative values slightly away from cyan.
The freed color space was utilized to represent the lower positive temperatures.
A gradient from white to cyan $0^{\circ}${C}-$10^{\circ}${C} is followed by a gradient from cyan to green to represent the moderate temperature range of $0^{\circ}${C}-$20^{\circ}${C}.
Next to this, a subsequent gradient from yellow through beige to light brown shows values between $20^{\circ}${C} and $40^{\circ}${C}. 
A further transition to dark red finally shows higher temperature range of up to $53^{\circ}${C}.

Our colormap semantics were designed to roughly differentiate between five temperature zones: very cold (blue to light blue), moderately cool (white to cyan), moderately warm (cyan to green), warm (green to yellow to beige) and hot (red).
Concerning red-green colorblind viewers, we used a lower and not overlapping lightness range for the red gradient and the green gradient.
The respective color gradients were separately optimized for local uniformity.
The panels~\textbf{2a} and \textbf{2b} of \autoref{fig:application_LittleBit} show the visualizations of the temperature data and the test function with the modified colormap.
Note that we used the \texttt{Little Bit} test function only for the upper part of the colormap that corresponds to temperature values between $m=5^{\circ}${C} and $M=53^{\circ}${C}.
As a result of our modifications of the colormap, it is now possible to see much more detail in the inhabited part of our planet and to distinguish between the different temperature zones.
Compared to~\textbf{1c}, the \texttt{Color Difference Field}~\textbf{2c} shows an increase in the color difference at the expense of the local uniformity of the positive data range.

%% file: 7_0_Conclustion.tex
%

\section{Conclusion} \label{section:conclusion}
In this paper, we have introduced the approach of using test functions as a standard evaluation method, and we have presented a test suite for continuous colormaps.
Like in other fields of computer science, one could use such test functions besides user-centered evaluation (e.g., user testimonies and empirical studies).
In compassion with user-centered evaluation, there is no need to recruit participants, design questionnaires or stimuli, organize payment, arrange experiment time and environment, and provide apparatus. Evaluating colormaps using the test suite can be conducted quickly and easily.
The designer can test many optional colormaps against many test functions and data sets, which is usually not feasible with user-centered evaluation. The same tests can be repeated with consistent control and comparability.

For the test suite, we first focused on the specific challenges of scalar fields.
The \autoref{subsubsection:neighbourhoodVariations}-\ref{subsubsection:littleBitVariations} describe the test functions we chose to address these challenges.
To help users with a less mathematical background, we tried to develop intuitive functions that are simple and easy to interpret.
The test suite currently includes step functions, different gradients, minima, maxima, saddle points, ridge and valley lines, global topology, thresholds, different frequencies, and a test for very small value changes.
Although these test functions cannot cover all possible challenges, we have laid down a solid foundation that can be extended continually.
We have also included the option to add noise to extend the possibilities of the basic test functions.  

Besides our newly designed functions, we have presented in \autoref{subsubsection:testcollection} a collection of functions used for evaluation in other computer science fields.
We think they will prove to be useful for the evaluation of colormaps as well.
Furthermore, we have included an initial selection of real-world data sets from different application areas.
%
%
As described in \autoref{subsection:realWorldData}, tests against real-world data are important in practice. 
Each real-world data set in our test suite presents an individual challenge of a combination of in scalar field analysis. 
Here, our intention is to provide a broad cover such that users are less dependent on external data.

Our test suite has been integrated into the open-access CCC-Tool.
In \autoref{section:testevaluation} we describe means to evaluate the results of the test functions visually and numerically that we have also implemented into our online-tool. 
An example of using the test suite to evaluate and enhance a user-designed colormap concerning a specific application problem is finally presented and discussed in \autoref{section:application}.

For a long-term perspective, we plan to continue the extension of our collection.
One option for real world data would be an open source database with a web interface and a link to our tool.
In order to adopt the test suite as a standard evaluation method, we would like to work on the method of automatic test reports, which can perform automatic analysis of a colormap with a set of tests chosen by the user.
%